\documentclass[nonacm,sigplan]{acmart}

\usepackage{balance}
\usepackage{tikz}
\usetikzlibrary{quantikz2}
\usepackage{amsfonts}
\usepackage{pifont}
\usepackage{xspace}
\usepackage[capitalise]{cleveref}
\usepackage{subcaption}
\usepackage{tcolorbox}
\usepackage{algorithm}
\usepackage[noend]{algpseudocode}
\usepackage{numprint}
\usepackage{enumitem}
\setlist[description]{leftmargin=\parindent,labelindent=\parindent}

\npthousandsep{,}
\definecolor{aliceblue}{rgb}{0.97, 0.96, 1.0}

\newenvironment{mybox}[1][blue!8]{  %
    \begin{tcolorbox}[   %
        left=0pt,
        right=0pt,
        top=0pt,
        bottom=0pt,
        colback=#1,
        colframe=#1,
        width=0.99\dimexpr\columnwidth\relax,
        boxsep=2pt,
        arc=0pt,outer arc=0pt,
    ]
}{
    \end{tcolorbox}
}

\Crefname{figure}{Fig.}{Figs.}
\Crefname{tabular}{Tab.}{Tabs.}
\Crefname{section}{\S}{\S}
\Crefname{theorem}{Thm.}{Thms.}
\Crefname{lemma}{Lem.}{Lems.}
\Crefname{corollary}{Cor.}{Cors.}
\Crefname{algorithm}{Alg.}{Algs.}
\Crefname{example}{Ex.}{Exs.}
\Crefname{definition}{Def.}{Defs.}

\DeclareMathOperator*{\argmin}{argmin}

\newcommand{\algorithmiccontinue}{\textbf{continue}}
\newcommand{\Continue}{\State \algorithmiccontinue}

\renewcommand{\paragraph}[1]{\textbf{{\emph {#1}}.~~~}}

\newcommand{\ours}{\textsc{guoq}\xspace}
\newcommand{\ourscaps}{\textsc{GUOQ}\xspace}
\newcommand{\oursrr}{\textsc{guoq-rewrite}\xspace}
\newcommand{\oursresynth}{\textsc{guoq-resynth}\xspace}
\newcommand{\oursbeam}{\textsc{guoq-beam}\xspace}
\newcommand{\oursrrfirst}{\textsc{guoq-seq-rewrite-resynth}\xspace}
\newcommand{\oursresynthfirst}{\textsc{guoq-seq-resynth-rewrite}\xspace}

\newcommand{\queso}{\textsc{queso}\xspace}
\newcommand{\ibm}{\textsc{ibmq\oldstylenums{20}}\xspace}
\newcommand{\newibm}{\textsc{ibm-eagle}\xspace}
\newcommand{\cliffordt}{Clifford + $\tgate$\xspace}
\newcommand{\ion}{\textsc{ionq}\xspace}
\newcommand{\tket}{\textsc{tket}\xspace}
\newcommand{\voqc}{\textsc{voqc}\xspace}
\newcommand{\bqskit}{BQSKit\xspace}
\newcommand{\synthetiq}{Synthetiq\xspace}
\newcommand{\pyzx}{PyZX\xspace}

\newcommand{\cx}{\ensuremath{CX}\xspace}
\newcommand{\tgate}{\ensuremath{T}\xspace}
\newcommand{\hadamard}{\ensuremath{H}\xspace}
\newcommand{\rz}[1]{\ensuremath{R_z^{\theta_{#1}}}\xspace}
\newcommand{\rza}[1]{\ensuremath{R_z^{{#1}}}\xspace}

\newcommand{\qubit}{\ensuremath{q}}
\newcommand{\circuit}{\ensuremath{C}}
\newcommand{\circuittype}{\mathcal{C}}
\newcommand{\subcircuit}{\ensuremath{\circuit_s}}
\newcommand{\ggate}{\ensuremath{g}}

\newcommand{\unified}[1]{\ensuremath{\tau_{\epsilon_{#1}}}}
\newcommand{\unifiedrw}{\ensuremath{\tau_{0}}}
\newcommand{\unifiedset}{\ensuremath{\mathcal{T}}}

\newcommand{\node}{\ensuremath{v}}

\newcommand{\real}{\ensuremath{\mathbb{R}}}

\newcommand{\unitary}{\ensuremath{U}}
\newcommand{\hs}{\ensuremath{\Delta}}
\newcommand{\hslong}[2]{\ensuremath{\sqrt{1-\frac{\Vert Tr(#1^\dag #2)\Vert ^2}{N^2}}}}
\newcommand{\synthesize}{\textsc{resynth}}
\newcommand{\cost}{\textsc{cost}}
\newcommand{\approxx}[1]{\ensuremath{\equiv_{\epsilon_{#1}}}}
\newcommand{\error}{\mathrm{error}}

\newcommand{\nisq}{\textsc{nisq}\xspace}
\newcommand{\ftqc}{\textsc{ftqc}\xspace}
\newcommand{\cmark}{\color{black}{\ding{51}}}
\newcommand{\xmark}{\color{red}{\ding{55}}}

\begin{document}

\title{Optimizing Quantum Circuits, Fast and Slow}

\author{Amanda Xu}
\affiliation{%
  \institution{University of Wisconsin-Madison}%
  \city{Madison, WI}
  \country{USA}%
}
\email{axu44@wisc.edu}

\author{Abtin Molavi}
\affiliation{%
  \institution{University of Wisconsin-Madison}%
  \city{Madison, WI}
  \country{USA}%
}
\email{amolavi@wisc.edu}

\author{Swamit Tannu}
\affiliation{%
  \institution{University of Wisconsin-Madison}%
  \city{Madison, WI}
  \country{USA}%
}
\email{swamit@cs.wisc.edu}

\author{Aws Albarghouthi}
\affiliation{%
  \institution{University of Wisconsin-Madison}%
  \city{Madison, WI}
  \country{USA}%
}
\email{aws@cs.wisc.edu}

\begin{abstract}
Optimizing quantum circuits is critical: the number of quantum operations needs to be minimized for a successful evaluation of a circuit on a quantum processor.
In this paper we unify two disparate ideas for optimizing quantum circuits,
\emph{rewrite rules}, which are fast standard optimizer passes, and \emph{unitary synthesis},
which is slow, requiring a search through the space of circuits.
We present a clean, unifying framework for thinking of rewriting and resynthesis as abstract circuit transformations.
We then present a radically simple algorithm, \ours, for optimizing quantum circuits that
exploits the synergies of rewriting and resynthesis.
Our extensive evaluation demonstrates the ability of \ours to strongly outperform existing optimizers on a wide range of benchmarks. 
\end{abstract}

\maketitle %
\pagestyle{plain} %

\section{Introduction}
\begin{figure}[t]
    \centering
    \includegraphics[width=\linewidth]{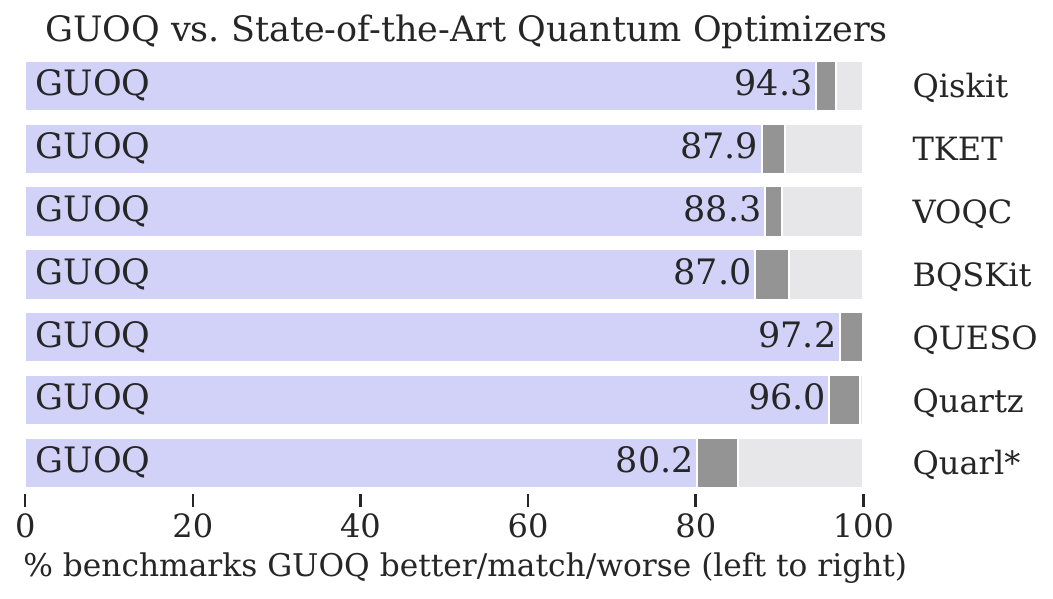}
    \caption{Summary of \ours compared to state-of-the-art on 2-qubit-gate reduction for the \ibm gate set.
    \ours and \bqskit are allowed to approximate the circuit up to $\epsilon = 10^{-8}$.
    *Quarl requires an NVIDIA A100 (40GB) GPU to run.}
    \label{fig:eval_summary}
\end{figure}
Quantum computing enables efficient simulation of quantum mechanical phenomena,
promising to catalyze advances in quantum physics, chemistry, materials science, and beyond.
Near-term quantum computers with more than a thousand
qubits operating in a noisy environment without error correction have already been deployed,
marking the current era of \emph{Noisy Intermediate Scale Quantum} (\nisq) computing~\cite{preskill2018nisq}. Recent groundbreaking 
experiments have implemented error-corrected \emph{logical} qubits and demonstrated potential for 
reducing \emph{logical error} \cite{bluvstein2024quera,da2024demonstration}.
Although many challenges remain, \emph{fault-tolerant quantum computing} (\ftqc) is on the horizon.

In both \nisq and \ftqc, reducing errors is a critical obstacle to overcome.
Every quantum operation has a probability of failure causing a quantum execution
to quickly devolve into random noise. The \nisq paradigm aims to mitigate these errors
in the absence of error correction primarily by reducing the number of operations.
However, error correction in \ftqc is not a panacea and introduces its own unique 
bottlenecks~\cite{campbell2017roadtoft,skoric2023decodingbacklog}, which
can render the error correction scheme useless if left untamed.
Especially in the near term, \ftqc architectures may face challenges in handling 
large circuit depths due to physical imperfections such as two-level system (TLS) drift, 
qubit leakage, high-energy particle strikes, and crosstalk~\cite{acharya2023google, bluvstein2024quera, mcewen2022particlestrike}. 
Therefore, it is of utmost importance to reduce the number of operations for \ftqc as well.

Current approaches tackling quantum circuit optimization primarily apply 
\emph{peephole optimization} using a fixed set of \emph{rewrite rules}.
Some tools use a small set of hand-crafted rules \cite{nam2018automated,hietala2021voqc,kissinger2020Pyzx}, while others automatically synthesize 
rules~\cite{xu2023queso,xu2022quartz}. 
The idea is 
to apply rewrite rules in a schedule, transforming subcircuits to semantically equivalent ones with fewer operations.
\emph{Rewrite rules are fast} to apply---match a pattern and rewrite it---but inherently only perform local optimizations.

An orthogonal line of work has been studying the problem of \emph{unitary synthesis}.
A unitary matrix precisely represents the semantics of a quantum program.
Some quantum algorithms are simple to state in the form of a unitary but nontrivial
to decompose into elementary operations that can be executed on hardware \cite{grover1996,feynman1982hamiltonian}.
Thus, a large body of work has focused on synthesizing quantum circuits that implement a given
unitary matrix \cite{davis2020qsearch,smith2023leap,younis2021qfast,rakyta2022squander,rakyta2022squandertwo,amy2014middle,tucci2005kak,paradis2024synthetiq,kang2023modular}. 
Recent works \cite{wu2022qgo,patel2021QUEST} have applied
these algorithms to optimize quantum circuits by partitioning large circuits into manageably-sized 
\emph{subcircuits} consisting of a few qubits at most and then \emph{resynthesizing} each subcircuit to produce a new 
subcircuit whose unitary is equivalent, or close enough, to the original subcircuit's unitary.
\emph{Unitary synthesis is slow}: usually a  combinatorial search problem through the space of circuits,
but can apply to deep subcircuits.

\begin{table}[t]
    \centering
    \begin{tabular}{lcc}
            \toprule
            & Rewrite rules & Resynthesis\\
            \midrule
            Fast & \cmark & \xmark \\
            Limited by \# gates & \cmark & \xmark \\
            Limited by \# qubits & \xmark & \cmark \\
            Approximate & \xmark & \cmark \\
            \bottomrule
    \end{tabular}
    \caption{Characteristics of rewrite rules and resynthesis.}
    \label{tbl:comparison}
\end{table}

Rewrite rules and resynthesis have their own strengths and weaknesses---see \cref{tbl:comparison}.
Individual rewrite rules are fast to apply but are limited to small patterns with few gates.
On the other hand, resynthesis is slow but can support circuits with an arbitrary number of gates
because the primary limiting factor is the number of qubits.
Critically, resynthesis
has the power to optimize deep subcircuits
and it can \emph{approximate} the solution to some degree. Rewrite rules are too small and rigid to find meaningful 
approximations. Approximations can unlock new circuit optimizations but must be applied carefully to avoid 
introducing too much error.

Inspired by how humans combine fast and slow modes of thinking~\cite{kahneman2011thinking}, \emph{Systems 1 and 2}, we ask the following question: 
\begin{center}
    \emph{
    Can we design an optimization approach that can synergistically combine the powers of optimizing quantum circuits fast, via rewrite rules, and slow, via resynthesis?
    }
\end{center}
We answer this question affirmatively: we demonstrate that we can unify rewriting and resynthesis and propose a simple optimization algorithm that significantly outperforms 
existing approaches in the literature.

\begin{figure}[t]
    \includegraphics[scale=1]{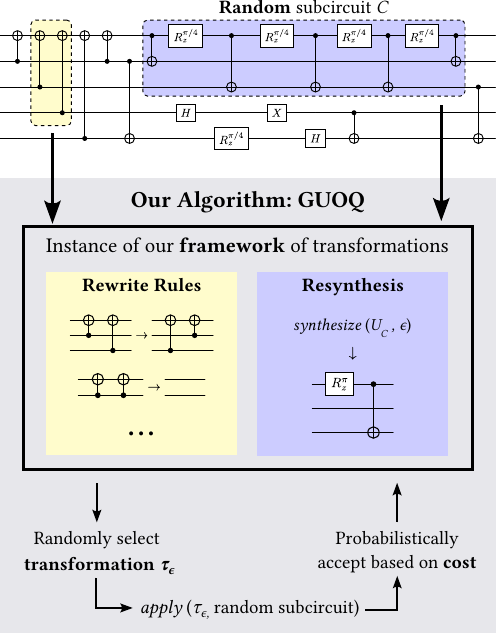}
    \caption{Overview of our approach}\label{fig:overview}
\end{figure}

\paragraph{Our approach} We propose a framework to unify rewrite rules and resynthesis for optimizing quantum circuits (see \cref{fig:overview}).
The key insight is that we can abstract both optimizations into a closed-box circuit \emph{transformation}
with a degree of approximation $\epsilon$. Our flexible and generic framework
allows arbitrary transformations, which can be applied freely. We prove a simple additive upper bound 
on the final approximation after applying a sequence of transformations.

We can instantiate our framework using a set of circuit transformations.
The key challenge is deciding in what order to apply these transformations---this is the \emph{phase ordering problem} that has plagued optimizing compilers for decades!
We discover that, perhaps surprisingly, a simple and lightweight 
\emph{simulated annealing}-like algorithm is the most effective solution, outperforming
more clever heuristics or algorithms. 
The lack of structure in our problem lends itself
to an approach that randomly and quickly alternates between fast and slow optimization, as opposed to sophisticated
approaches guided by hand-crafted heuristics or reinforcement learning.
Let this serve as another bitter lesson~\cite{bitterLesson} 
that simple methods often prevail. 

We implement our algorithm, \ours (Good Unified Optimizations for Quantum), and provide an extensive evaluation
against state-of-the-art optimizers and \emph{superoptimizers} using a benchmark suite with 247 diverse quantum circuits
implementing near- and long-term algorithms. 
Our evaluation demonstrates the following: (1)  \ours significantly outperforms state-of-the-art tools (see \cref{fig:eval_summary} for a summary),
(2) \ours's randomized search approach is critical for efficiently combining rewriting and resynthesis,
and (3) \ours can flexibly perform well in the \nisq and \ftqc regimes.
For instance, \ours outperforms the recent superoptimizer, Quarl~\cite{li2024quarl}, in terms of 2-qubit-gate reduction
on 80\% of the benchmarks. This is despite the fact that Quarl uses a specialized deep reinforcement learning technique for quantum-circuit optimization, requires more computational resources (GPUs),
and has been trained on portions of the benchmark suite. 

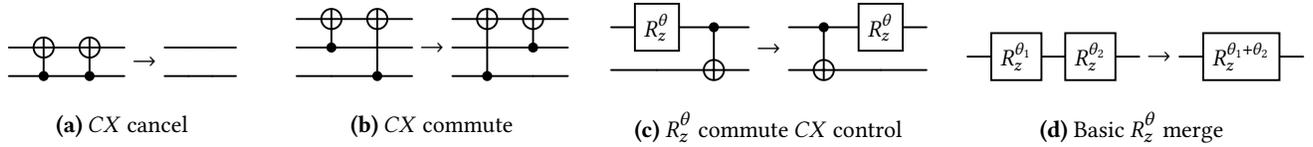
\begin{figure*}[t!]
    \begin{subfigure}[t]{0.22\linewidth}
        \centering
        \begin{tikzpicture}
            \node[scale=0.9] {
                \begin{quantikz}[align equals at=1.5, row sep = {12pt,between origins}, column sep = 10pt]
                    & \targ{}   & \targ{}   & \\
                    & \ctrl{-1} & \ctrl{-1} & 
                \end{quantikz}$\to$\begin{quantikz}[align equals at=1.5, row sep = {12pt,between origins}, column sep = 10pt]
                    &&& \\
                    &&&
                \end{quantikz}
            };
        \end{tikzpicture} 
        \caption{$\cx$ cancel}
        \label{fig:cx-cancel}
    \end{subfigure}
    \hfill
    \begin{subfigure}[t]{0.23\linewidth}
        \centering
        \begin{tikzpicture}
            \node[scale=0.9] {
                \begin{quantikz}[align equals at=2, row sep = {12pt,between origins}, column sep = 10pt]
                    & \targ{}   & \targ{}   & \\
                    & \ctrl{-1} &           & \\
                    &           & \ctrl{-2} & 
                \end{quantikz}$\to$\begin{quantikz}[align equals at=2, row sep = {12pt,between origins}, column sep = 10pt]
                    & \targ{}   & \targ{}   & \\
                    &           & \ctrl{-1} & \\
                    & \ctrl{-2} &           & 
                \end{quantikz}
            };
        \end{tikzpicture} 
        \caption{$\cx$ commute}
        \label{fig:cx-commute}
    \end{subfigure}
    \hfill
    \begin{subfigure}[t]{0.26\linewidth}
        \centering
        \begin{tikzpicture}
            \node[scale=0.9] {
                \begin{quantikz}[align equals at=1.5, row sep = {18pt,between origins}, column sep = 10pt]
                    & \gate{\rz{}} & \ctrl{1} & \\
                    &              & \targ{}  & 
                \end{quantikz}$\to$\begin{quantikz}[align equals at=1.5, row sep = {18pt,between origins}, column sep = 10pt]
                    & \ctrl{1} & \gate{\rz{}} & \\
                    & \targ{}  &                   & 
                \end{quantikz}
            };
        \end{tikzpicture} 
        \caption{$\rz{}$ commute $\cx$ control}
        \label{fig:rz-commute-cx}
    \end{subfigure}
    \hfill
    \begin{subfigure}[t]{0.27\linewidth}
        \centering
        \begin{tikzpicture}
            \node[scale=0.9] {
                \begin{quantikz}[align equals at=1, row sep = {18pt,between origins}, column sep = 10pt]
                    & \gate{\rz{1}} &  \gate{\rz{2}} & 
                \end{quantikz}$\to$\begin{quantikz}[align equals at=1, row sep = {18pt,between origins}, column sep = 10pt]
                    & \gate{\rza{\theta_1 + \theta_2}} & 
                \end{quantikz}
            };
        \end{tikzpicture} 
        \caption{Basic $\rz{}$ merge}
        \label{fig:rz-merge}
    \end{subfigure}
    \caption{Examples of rewrite rules. Observe how the rules with $\rz{}$ use \emph{symbolic} $\theta$ angles.}
    \label{fig:rewrite-rules}
\end{figure*}

\paragraph{Contributions} We make the following contributions:
\begin{itemize}
    \item A framework that abstracts the inner workings of rewriting and resynthesis 
    into closed-box circuit \emph{transformations} with \emph{approximate} semantic guarantees.
    \item A lightweight algorithm, \ours, inspired by simulated annealing, that searches the space of transformations.
    \item An implementation of \ours and thorough 
    evaluation considering both \nisq and \ftqc that demonstrates its effectiveness 
    compared to state-of-the-art optimizers. 
\end{itemize}

\section{Background and Overview}
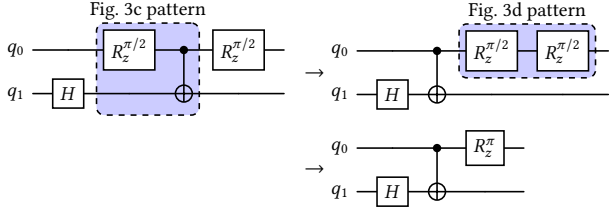
\begin{figure}[t]
    \centering
    \begin{tikzpicture}
        \node[scale=0.75] (first) {
            \begin{quantikz}[align equals at=1.5, row sep = {22pt,between origins}, column sep = 10pt]
                \lstick{$\qubit_0$}    & & \gate{\rza{\pi/2}}\gategroup[2,steps=2,style={rounded corners, dashed, inner sep=0pt, fill=blue!20}, background]{{\cref{fig:rz-commute-cx} pattern}} & \ctrl{1} & \gate{\rza{\pi/2}} &\\
                \lstick{$\qubit_1$}    & \gate{H}     &      & \targ{}  &                    & 
            \end{quantikz}
        };
        \node[scale=0.75, right = 0pt of first] (second) {    
            $\to$\begin{quantikz}[align equals at=1.5, row sep = {22pt,between origins}, column sep = 10pt]
                \lstick{$\qubit_0$}    &          & \ctrl{1} & \gate{\rza{\pi/2}}\gategroup[1,steps=2,style={rounded corners, dashed, inner sep=0pt, fill=blue!20}, background]{{\cref{fig:rz-merge} pattern}}  & \gate{\rza{\pi/2}} &\\
                \lstick{$\qubit_1$}    & \gate{H} & \targ{}  &                    &                    & 
            \end{quantikz}
        };
        \node[scale=0.75, below = 1.5cm of second.west, anchor=west] {
            $\to$\begin{quantikz}[row sep = {22pt,between origins}, column sep = 10pt]
                \lstick{$\qubit_0$}    &          & \ctrl{1} & \gate{\rza{\pi}} &\\
                \lstick{$\qubit_1$}    & \gate{H} & \targ{}  &                    & 
            \end{quantikz}
        };
    \end{tikzpicture}
    \caption{Example of applying the rule from \cref{fig:rz-commute-cx} followed by the rule from \cref{fig:rz-merge}.}
    \label{fig:rewrite-rule-ex}
\end{figure} 
In this section, we provide the high-level background required for understanding 
our approach and use examples to highlight the differences between rewrite rules and resynthesis. 
We also describe an overview of our approach along with some 
concrete examples showing how rewrite rules and resynthesis can work together.

\subsection{Quantum Circuits Background}

\paragraph{Quantum circuits} 
Quantum circuits are composed of combinations 
of quantum operations (or \emph{gates}) applied to \emph{qubits}.
Some operations, like the \emph{Hadamard} gate ($\hadamard$), are only applied
to a single qubit, whereas operations like the \emph{controlled-NOT} gate ($\cx$)
apply to an ordered pair of qubits. 
Another common class of quantum gates includes 
rotational gates parameterized on input angles, such as the $\rz{}$ gate. 

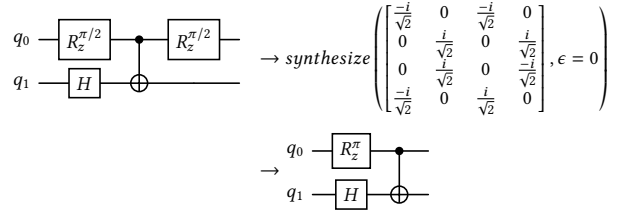
\begin{figure}[t]
    \centering
    \begin{tikzpicture}
        \node[scale=0.75] (first) {
            \begin{quantikz}[align equals at=1.5, row sep = {22pt,between origins}, column sep = 10pt]
            \lstick{$\qubit_0$}    & \gate{\rza{\pi/2}} & \ctrl{1} & \gate{\rza{\pi/2}} &\\
            \lstick{$\qubit_1$}    & \gate{H}           & \targ{}  &                    & 
            \end{quantikz}
        };
        \node[scale=0.75, right = 0pt of first] (second) {
            $\to synthesize \left( \begin{bmatrix}
                \frac{-i}{\sqrt{2}} & 0 & \frac{-i}{\sqrt{2}} & 0\\
                0 & \frac{i}{\sqrt{2}} & 0 & \frac{i}{\sqrt{2}}\\
                0 & \frac{i}{\sqrt{2}} & 0 & \frac{-i}{\sqrt{2}}\\
                \frac{-i}{\sqrt{2}} & 0 & \frac{i}{\sqrt{2}} & 0\\
                \end{bmatrix}, \epsilon=0 \right) $
        };
        \node[scale=0.75, below = 1.5cm of second.west, anchor=west] {
            $\to$\begin{quantikz}[row sep = {22pt,between origins}, column sep = 10pt]
                \lstick{$\qubit_0$}    & \gate{\rza{\pi}} & \ctrl{1} &\\
                \lstick{$\qubit_1$}    & \gate{H}         & \targ{}  & 
            \end{quantikz}
        };
    \end{tikzpicture}
    \caption{Example of resynthesizing the initial \cref{fig:rewrite-rule-ex} circuit. }
    \label{fig:resynth-ex}
\end{figure}

Consider the example circuit on the left sides of \cref{fig:rewrite-rule-ex,fig:resynth-ex},
where each horizontal wire corresponds
to a qubit---e.g., the 
$\hadamard$ gate is the first gate applied to qubit $\qubit_1$.

\paragraph{Rewrite rules} A rewrite rule is a pair of semantically equivalent circuits. Although rewrite 
rules are in principle bidirectional, we will refer to the left-hand side as 
the \emph{pattern} and the right-hand side as the \emph{replacement} for simplicity.
\cref{fig:rewrite-rules} shows examples of some commonly used rewrite rules.
Applying rewrite rules to a circuit is simple. Begin by searching for 
a \emph{match} for the pattern and if one exists, substitute it with the replacement.
For example, in \cref{fig:rewrite-rule-ex}, there is a match for the pattern 
of the rule in \cref{fig:rz-commute-cx}, shown by the highlighted subcircuit.
Rewriting the match to the replacement allows the rule in \cref{fig:rz-merge} to apply
and reduces the gate count by one. This general idea of composing rewrite rules
is the heart of how we optimize quantum circuits using rewrite rules.

\paragraph{Resynthesis}
Circuit \emph{resynthesis} takes advantage of the vast line of work done in 
\emph{unitary synthesis} to resynthesize circuits according to some optimization
objective. A circuit can be represented as a unitary matrix.
Given a circuit's unitary matrix,
unitary synthesis constructs a new circuit, with fewer gates, whose unitary is within $\epsilon$ of the original unitary 
according to some distance metric.
Note that the original circuit's structure is lost by converting it into a unitary and 
the synthesis algorithm needs to search for a new circuit structure from scratch.
This is inherently a slow search through the space of circuits, e.g., the \bqskit compiler \cite{younis2021bqskit}
performs a bottom-up search using two-qubit subcircuits. 

\cref{fig:resynth-ex} illustrates circuit resynthesis using the same initial circuit as
\cref{fig:rewrite-rule-ex} and no approximation by setting $\epsilon=0$.
Observe how the 
final circuit is equivalent to the result from applying
rewrite rules because we can apply the rule in \cref{fig:rz-commute-cx} to push 
the \rza{\pi} gate across the control of the $\cx$.

\subsection{Comparing Rewriting and Resynthesis}
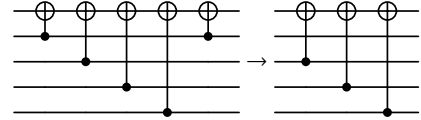
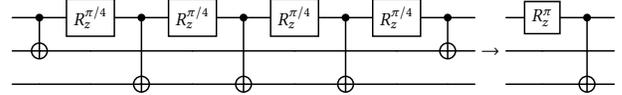
\begin{figure}[t]
    \begin{subfigure}{\linewidth}
        \centering
        \begin{tikzpicture}
            \node[scale=0.8] {
                \begin{quantikz}[align equals at=3, row sep = {12pt,between origins}, column sep = 10pt]
                    & \targ{}   & \targ{}   & \targ{}   & \targ{}   & \targ{}   & \\
                    & \ctrl{-1} &           &           &           & \ctrl{-1} & \\
                    &           & \ctrl{-2} &           &           &           & \\
                    &           &           & \ctrl{-3} &           &           & \\
                    &           &           &           & \ctrl{-4} &           & 
                \end{quantikz}$\to$\begin{quantikz}[align equals at=3, row sep = {12pt,between origins}, column sep = 10pt]
                    & \targ{}   & \targ{}   & \targ{}   & \\
                    &           &           &           & \\
                    & \ctrl{-2} &           &           & \\
                    &           & \ctrl{-3} &           & \\
                    &           &           & \ctrl{-4} & 
                \end{quantikz}
            };
        \end{tikzpicture} 
        \caption{Rewrite rules better}
        \label{rewrite-good}
    \end{subfigure}
    \begin{subfigure}{\linewidth}
        \centering
        \begin{tikzpicture}
            \node[scale=0.7] {
                \begin{quantikz}[align equals at=2, row sep = {18pt,between origins}, column sep = 10pt]
                    & \ctrl{1} & \gate{\rza{\pi/4}} & \ctrl{2} & \gate{\rza{\pi/4}} & \ctrl{2} & \gate{\rza{\pi/4}} & \ctrl{2} & \gate{\rza{\pi/4}} & \ctrl{1} &  \\
                    & \targ{}&  &  &  &  &  &  &  & \targ{}&                                                                                   \\
                    &  &  & \targ{}&  & \targ{}&  & \targ{}&  &  &                                                                                 
                \end{quantikz}$\to$\begin{quantikz}[align equals at=2, row sep = {18pt,between origins}, column sep = 10pt]
                    & \gate{\rza{\pi}} & \ctrl{2} & \\
                    &                  &          & \\
                    &                  & \targ{}  &
                \end{quantikz}
            };
        \end{tikzpicture} 
        \caption{Resynthesis better}
        \label{resynth-good}
    \end{subfigure}
    \caption{Comparing rewrite rules and resynthesis.}
\end{figure}

Recall that resynthesis is limited by the number of qubits in the circuit, because the size of the unitary is exponential in the number of qubits.
\cref*{rewrite-good} is an example of a circuit where it is better to apply rewrite rules.
The structure closely resembles the circuit for the \emph{quantum Fourier transform} (QFT)~\cite{coppersmith2002qft},
a critical subroutine in many quantum algorithms.
This circuit involves too many qubits for unitary synthesis to succeed in a reasonable
amount of time. 
However, it only takes a few applications of two rewrite rules (\cref{fig:cx-cancel,fig:cx-commute}) to 
quickly get to the right-hand side. 
Even if we partition this circuit into more tractable 3-qubit subcircuits to resynthesize, 
we would not be guaranteed to reach the right-hand side. In fact, it would require a 
series of lucky coincidences over multiple rounds of partitioning the circuit.

Resynthesis involving fewer qubits though can compensate for the limited-sized patterns 
in rewrite rules.
\cref*{resynth-good} is an example of a deep circuit where resynthesis can accelerate the search. Although we can achieve the 
optimized circuit using rewrite rules, it requires a long 
sequence of several of the rules from \cref{fig:rewrite-rules} in a very specific order.
Resynthesis can discover the circuit on the right-hand side all at once because the 
circuit only involves 3 qubits.

As we saw in the above examples, there are complementary qualities between rewrite rules and resynthesis.

\subsection{Our Approach}

\paragraph{Unifying rewrite rules and resynthesis}
Our framework to unify rewrite rules and resynthesis introduces an abstraction
for \emph{transforming} circuits.
We specify a function called a circuit \emph{transformation}, denoted $\unified{}$,
that returns a circuit that is semantically equivalent to the original up to the approximation $\epsilon$.
Beyond this guarantee, transformations are closed-box. 
Given a set of rewrite rules and resynthesis algorithms, we can instantiate a set 
of transformations. 
Crucially, our framework allows us to apply circuit transformations in any order, 
and we prove an upper bound on 
the final approximation degree
when applying an arbitrary sequence of transformations.

\paragraph{Optimization objectives} Optimizing quantum circuits requires 
diverse optimization objectives depending on the application. In \nisq, two-qubit gates
are the dominant source of noise whereas in \ftqc, $\tgate$ gates are the most costly
to perform in an error-corrected fashion, followed by two-qubit gates.
We view these gate-minimizing optimization objectives as \emph{soft} constraints in our search. 
Our \emph{hard} constraint when resynthesizing subcircuits is staying within the specified
global error threshold $\epsilon_f$. Allowing more error can allow the synthesis algorithm to find
a solution with fewer gates, so it is critical to find a balance between these two
competing optimization objectives. For example, an objective for \nisq might be the following:
$\argmin_{\circuit'} \ \textsc{2q-count}(\circuit') \ \text{s.t.\ } \epsilon_{\circuit'} \leq \epsilon_f$.

\paragraph{The \ours algorithm}
The vast and discrete landscape for optimizing quantum circuits
provides sparse navigation for traversing it. 
\ours is our simple algorithm inspired by simulated annealing~\cite{kirkpatrick1983simulatedannealing}
that rapidly and randomly searches the space of transformations.
We find that an approach like \ours is well-suited for solving our problem 
because it has minimal memory requirements, is easy to implement,
and explores the solution space much faster than other approaches. 
At a high level, the algorithm maintains a single candidate and applies randomly chosen transformations
 to randomly chosen subcircuits. Transformations with $\epsilon = 0$ can be applied an unlimited number of times
while transformations with $\epsilon > 0$ are limited based on the target error threshold.
If a transformation preserves or reduces gates related to the optimization objective,
it is always accepted. Otherwise, it is only accepted with a small probability.

\begin{figure}[t]
    \includegraphics[width=0.49\linewidth]{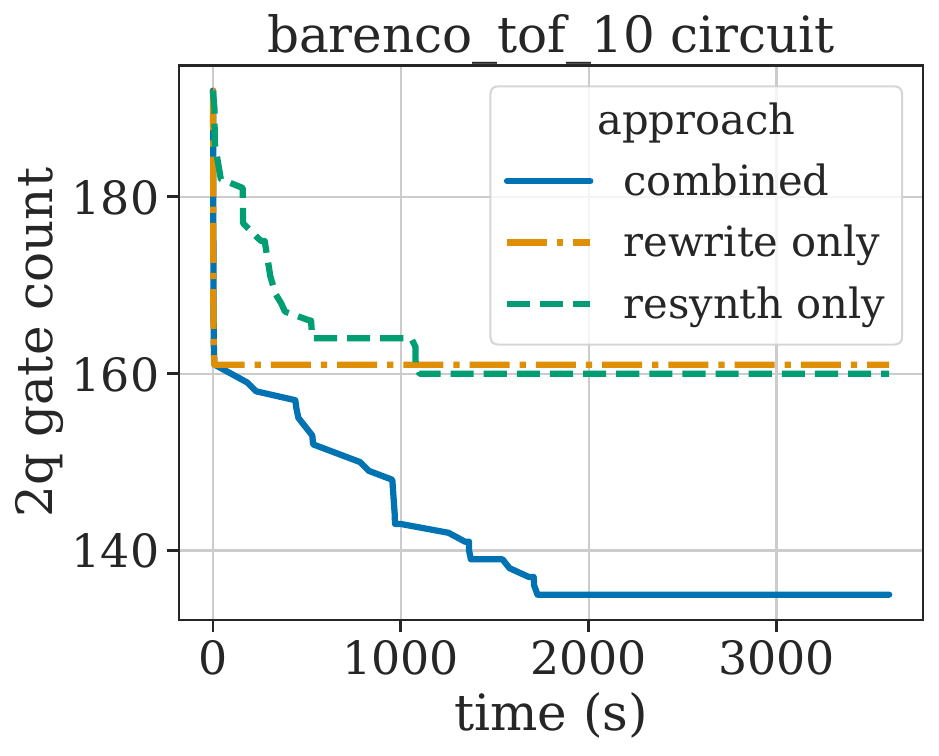}
    \includegraphics[width=0.49\linewidth]{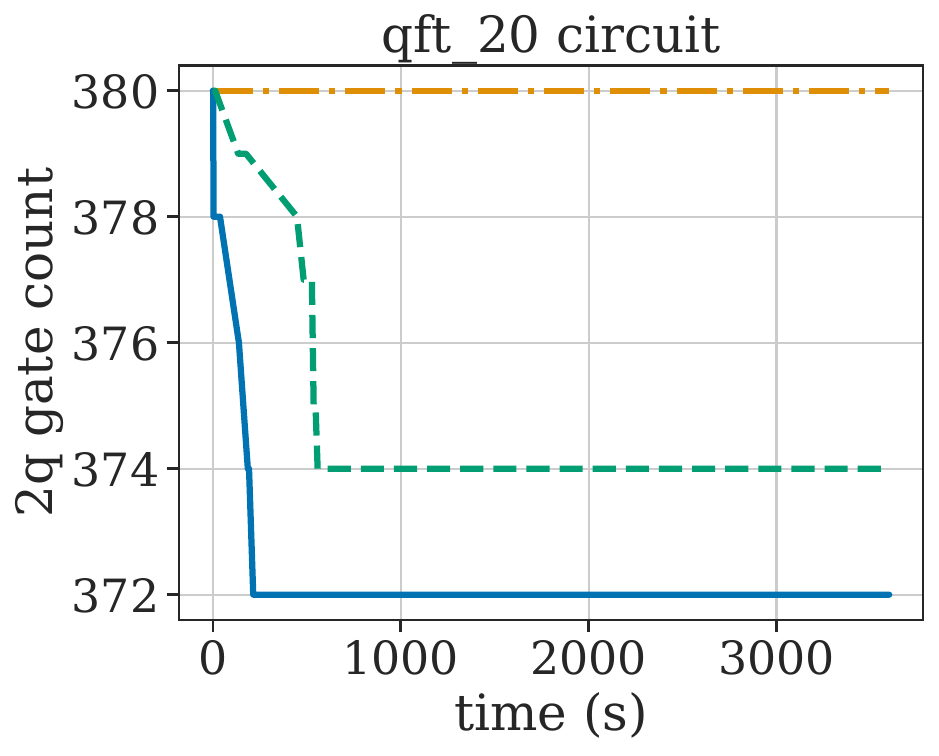}
    \caption{An example showing the two-qubit gate count of the current best solution
    for \texttt{barenco\_tof\_10} and \texttt{qft\_20}
    over an hour of search using 1) only rewrite rules, 2) only resynthesis, and 
    3) rewrite rules and resynthesis combined.}
    \label{fig:overview_anecdote}
\end{figure}

\paragraph{A concrete example} As a primer, we provide an example demonstrating the benefits 
of combining rewrite rules and resynthesis on the \texttt{barenco\_tof\_10} benchmark, 
which is an implementation of a multi-control Toffoli gate~\cite{barenco1995toff}, and the 
\texttt{qft\_20} benchmark, which implements the \emph{quantum Fourier transform} (QFT)~\cite{coppersmith2002qft}.
These benchmarks are critical building blocks for the famous Shor's algorithm~\cite{shor1994}.
\cref{fig:overview_anecdote} shows the two-qubit gate count of the best solution
over an hour of search for \ours using 1) rewrite rules only, 2) resynthesis only, and 3) 
rewrite rules and resynthesis combined. 

As we can see, rewrite rules on their own
quickly get stuck in a local minimum for almost the entire search, whereas 
resynthesis on its own is able to gradually make progress but is too slow.
Combining rewrite rules and resynthesis allows resynthesis to progress the search
when rewrite rules get stuck by mutating the circuit and ``teleporting'' 
the candidate solution to a different area of the optimization landscape. 
Working in tandem, rewrite rules and resynthesis can go beyond the capabilities of either alone.

\section{Quantum Circuits and Approximations}
We now formalize the necessary background
for understanding quantum circuits and their semantics.
The notion of approximate semantics and subcircuits will 
be critical for our framework.

\paragraph{Semantics} A quantum gate is a linear transformation of the state vector.
A gate $\ggate$ acting on $m$ qubits 
can be represented by a $2^m\times2^m$ \emph{unitary} matrix $\unitary_\ggate$. 
Composing the gates in a circuit
with $n$ qubits using matrix multiplication results in a $2^n\times2^n$ unitary matrix
exactly representing the semantics of the circuit.

\begin{example}
    The semantics of the $\tgate$ and $\cx$ gates are the following unitary matrices:
    $$\unitary_\tgate \coloneq \begin{bmatrix}
        1 & 0\\
        0 & e^{i\pi/4}
    \end{bmatrix},~ \unitary_{\cx} \coloneq \begin{bmatrix}
        1 & 0 & 0 & 0\\
        0 & 1 & 0 & 0\\
        0 & 0 & 0 & 1\\
        0 & 0 & 1 & 0
    \end{bmatrix}$$

    Consider the circuit $\circuit \coloneq \tgate~ \qubit_1;~\cx~\qubit_0 ~ \qubit_1;$.
    The unitary $\unitary_\circuit$ representing the semantics of $\circuit$ 
    is precisely $\unitary_{\cx} (I\otimes \unitary_\tgate)$,
    where the tensor product $(I\otimes \unitary_\tgate)$ with the identity matrix denotes that the $\tgate$ gate is applied to the second qubit.
\end{example}

\paragraph{Circuit equivalence} Circuits are semantically equivalent if 
their unitaries are exactly equal. Additionally, some circuits may be equivalent 
up to \emph{global phase} because two quantum states $\ket{\psi_1}$ and $\ket{\psi_2} = e^{i\phi}\ket{\psi_1}$ 
are observationally indistinguishable \cite{nielsen2002quantum} for any angle $\phi\in \real$. 
Two circuits $\circuit$ and $\circuit'$ 
are semantically equivalent modulo global phase, denoted $\circuit \equiv \circuit'$,
if and only if $\unitary_\circuit = e^{i\phi}\unitary_{\circuit'}$ for some angle $\phi$.

\paragraph{Approximating circuits} We can approximate the semantics of a circuit $\circuit$
arbitrarily by defining a notion of \emph{distance}, which is a function of 
the original and approximated circuits' unitaries. The \emph{Hilbert--Schmidt distance} ($\hs$) 
is a convenient distance function
used in prior work \cite{patel2021QUEST, paradis2024synthetiq} due to its ability to handle equivalence modulo global phase
and ease of computation. The Hilbert--Schmidt distance between two unitary matrices is formally defined in 
\cref{def:hs}. Using this definition, we can define approximate circuit equivalence
as shown in \cref{def:approx-equiv}.

\begin{definition}[Hilbert--Schmidt distance]\label{def:hs}
    Let $\unitary$ and $\unitary'$ be two $N\times N$ unitary matrices.
    $\hs(\unitary, \unitary') \coloneq \hslong{\unitary}{\unitary'}$.
\end{definition}

\begin{definition}[Approximate circuit equivalence]\label{def:approx-equiv}
    Two circuits $\circuit$ and $\circuit'$ 
are $\epsilon$-equivalent, denoted $\circuit \approxx{} \circuit'$,
if and only if $\hs(\unitary_\circuit, \unitary_{\circuit'}) \leq \epsilon$.
\end{definition}

\paragraph{Subcircuits} 
To define a subcircuit, it is best to consider a DAG representation of a circuit, where the nodes are gates and the wires between gates 
are directed from left to right. 
Then a subcircuit is precisely a \emph{convex}
subgraph of this DAG. As defined in prior work \cite{xu2022quartz},
a \emph{convex} subgraph is a subgraph that contains every path that exists between its nodes
in the original graph. Intuitively, this requirement enforces continuous qubit wires in the subcircuit.
For example, \cref{fig:overview} (top) shows two highlighted subcircuits.

\section{A Unified Optimization Framework}
\label{sec:framework}
We are ready to formally describe our framework unifying rewrite rules and resynthesis.
Our framework introduces abstract \emph{transformations} with an associated error 
tolerance $\epsilon$. We show how to represent rewrite rules and resynthesis 
in this unifying framework and prove an upper bound on the error when 
composing these transformations in arbitrary sequences. This
underlies the design of \ours, which applies transformations 
in arbitrary orders.

\subsection{Circuit Transformations}
The following definition presents an abstraction of 
a circuit transformation as a function that takes a circuit $\circuit$ and 
produces an $\epsilon$-equivalent $\circuit'$.

\begin{definition}[Transformation]\label{def:unified}
    Let $\circuittype$ represent a circuit type. 
    A transformation $\unified{} : \circuittype \to \circuittype$ 
    accepts a circuit $\circuit$ and returns a circuit $\circuit'$ such that
    $\circuit \approxx{} \circuit'$.
\end{definition}

Rewrite rules and resynthesis can be represented in this framework as transformations 
over subcircuits. Consider a rewrite rule $C_1 \to C_2$. 
The transformation $$ \unifiedrw(C) \coloneq \left\{ \begin{matrix}
    C_2 & \text{if }C = C_1 \\
    C & \text{otherwise}
  \end{matrix}\right.\, $$ captures the rewrite rule by transforming circuits that are syntactically equivalent to $C_1$ (up to qubit renaming)
 and acting as the identity function otherwise. Observe 
 how this transformation carries an $\epsilon = 0$ because rewrite rules 
preserve exact semantic equivalence.

Similarly, we can define a transformation representing resynthesis. Assume 
there exists a circuit resynthesis function $\synthesize: (\circuittype \times \real) \to \circuittype$ 
that given a circuit $\circuit$ and error tolerance $\epsilon$, returns a circuit 
$\circuit'$ such that $\circuit \approxx{} \circuit'$.
The transformation is simply $\unified{}(C) \coloneq \synthesize(C, \epsilon)$.
An example of such a circuit resynthesis function is a thin wrapper around 
a unitary synthesis function that computes the input circuit's unitary 
before invoking unitary synthesis.

\subsection{Composing Transformations}
Composing transformations is not as straightforward as composing rewrite rules
because transformations are allowed to approximate the circuit. Prior work \cite{patel2021QUEST}
shows how to upper bound the error when approximating \emph{disjoint} partitions 
of a circuit. 
We present a flexible and generic framework 
that allows us to apply transformations in an arbitrary fashion.
In particular, we can apply a transformation to a subcircuit 
that only \emph{partially} contains a previously transformed subcircuit.

Formalized in \cref{theorem:error_bound}, we prove that the upper bound on the 
error when composing a finite sequence of transformations
is the sum of all the errors from each transformation.
Without loss of generality, we can assume all transformations
have the same error.

\begin{theorem}\label{theorem:error_bound}
Suppose we are given a set of transformations $\unified{}^1, \ldots, \unified{}^n$.
    Let $\circuit_0,\ldots,\circuit_n$ be a sequence of circuits such that 
    $\circuit_i$ is the result of applying transformation $\unified{}^i$ to a subcircuit 
    of $\circuit_{i-1}$ for all $1 \leq i \leq n$. 
    Then, $\circuit_0 \equiv_{n\epsilon} \circuit_n$.
\end{theorem}

\section{\ourscaps: A Stochastic Algorithm}
\label{sec:algorithm}
In this section, we begin by formally stating the quantum-circuit optimization problem.
Optimizing quantum circuits is hard because of the large search space and the 
difficulty of simulating quantum circuits.
Next we describe our algorithm $\ours$ for solving this problem given an 
instantiation of our framework. $\ours$ is fast, flexible, and easy to implement.
It applies a given set of transformations in a randomized fashion inspired by \emph{simulated annealing}, a classic algorithm for solving discrete optimization problems. Finally, we discuss implementation details
for optimizing our algorithm.

\subsection{Optimization objectives for quantum circuits}

Different quantum computing hardwares and paradigms will require 
unique optimization objectives. For example, on \nisq hardware 
it is critical to reduce two-qubit gate count. Other optimization
objectives include $\tgate$ count and circuit \emph{depth}, rather than gate count.
Our approach is flexible and we can define any cost function, $\cost : \circuittype \to \real$, to minimize, where $\circuittype$ is the set of all circuits.

\begin{example}[Multiple optimization objectives]
    Consider the \ftqc setting where we want to reduce primarily $\tgate$ gates,
    followed by $\cx$ gates. We can model this optimization objective by
    defining $\cost$ as
    $2 \cdot \#_\tgate(\circuit) + \#_{\cx}(\circuit)$,
    where  $\#_\tgate(\circuit)$ and $\#_{\cx}(\circuit)$ are the $\tgate$
    and $\cx$ gate counts, respectively.
\end{example}

Transformations in our framework can be approximate, so it is natural to accept as input
an error tolerance that the result should not exceed. 
Using this error tolerance as a hard constraint and $\cost$ as a soft 
constraint we can formulate the problem as a 
succinct constrained optimization problem.

\begin{definition}[Quantum-Circuit Optimization Problem]\label{def:optimize} 
    Given a circuit $\circuit$ and an \emph{error tolerance} $\epsilon_f \geq 0$, 
    the quantum-circuit optimization problem is the following
    constrained optimization problem:
    \begin{align*}
        & \argmin_{\circuit'} \cost(\circuit') \ \ \ \ \  \text{s.t.\ }  \hs(\circuit, \circuit') \leq \epsilon_f
    \end{align*}
\end{definition}

\subsection{The \ourscaps Algorithm}

We propose an algorithm inspired by simulated annealing. Simulated annealing is a general algorithm
for solving optimization problems with large search spaces. It has many nice properties
such as being fast, memory-efficient, easy to implement, and interruptible at any time to obtain a partial solution.
These properties, inherited in our algorithm, unlock an effective approach 
for solving a problem that is incredibly difficult to craft or learn predictive heuristics for.

\cref{fig:algs} shows the pseudocode for our algorithm. The inputs to the 
algorithm are the inputs to the quantum-circuit optimization problem and
a set of transformations $\unifiedset$. Given a set of rewrite rules
and resynthesis methods, we can instantiate $\unifiedset$ as discussed 
in \cref{sec:framework}. The ``moves'' we are allowed to make to modify 
the current solution are precisely the transformations in $\unifiedset$. 
The heart of the algorithm simply randomly samples
a transformation and randomly samples a subcircuit of the current solution circuit
to apply the transformation to. This move is always accepted if it
improves or preserves the quality of the solution with respect to $\cost$. 
Otherwise, it is accepted with some small probability that can be tuned 
using the temperature hyperparameter $t$. We adapt the standard acceptance
probability for simulated annealing~\cite{kirkpatrick1983simulatedannealing}, which approaches 0 as the 
candidate solution cost increases.

The remainder of the algorithm ensures the upper bound on the error in the 
final solution does not exceed the specified tolerance $\epsilon_f$.
Using \cref{theorem:error_bound}, we can simply keep track of the 
sum of all the errors across all transformations applied.
If applying a transformation would exceed the error bound,
then we abstain and skip to the next iteration where we have the opportunity
to sample a rewrite rule transformation with $\epsilon = 0$.
\cref{theorem:correct} states the correctness of \ours.

\begin{theorem}[Correctness of \ours]\label{theorem:correct}
    Let $\circuit'$ be the result of $\ours(\circuit, \epsilon_f, \unifiedset)$
    for any circuit $\circuit$, error tolerance $\epsilon_f \geq 0$, and set of transformations
    $\unifiedset$.
    Then, $\circuit \approxx{f} \circuit'$.
\end{theorem}

\begin{algorithm}[t]
  
    \caption{The \ourscaps Algorithm}\label{fig:algs}%
      \begin{algorithmic}[1]
      \Procedure{\ours}{circuit $C$, error $\epsilon_f$, transformations $\unifiedset$}
      \State initialize $C_\emph{best}$ and $C_\emph{curr}$ to $C$
      \State initialize $\error_\emph{best}$ and $\error_\emph{curr}$ to 0 
      \While {within time limit}
        \State {\color{ACMDarkBlue}Randomly select} transformation $\unified{}$ in  $\unifiedset$
        \If {$\error_\emph{curr} + \epsilon > \epsilon_{f}$}
            \Continue 
        \EndIf
        \State {\color{ACMDarkBlue}Randomly select} subcircuit $\subcircuit$ in $C_\emph{curr}$ 
        \State $C_\emph{curr}^\tau \gets$ result of replacing $\subcircuit$ with $\unified{}(\subcircuit)$ in $C_\emph{curr}$

        \If {$\cost(C_\emph{curr}^\tau) \leq \cost(C_\emph{curr})$}
          \State $C_\emph{curr} \gets C_\emph{curr}^\tau$
          \State $\error_\emph{curr} \gets \error_\emph{curr} + \epsilon$
        \Else ~ {\color{ACMDarkBlue}with probability} $\exp\left({-t\frac{\cost(C_\emph{curr}^\tau)}{\cost(C_\emph{curr})}}\right)$
          \State $C_\emph{curr} \gets C_\emph{curr}^\tau$
          \State $\error_\emph{curr} \gets \error_\emph{curr} + \epsilon$
        \EndIf
        \If {$\cost(C_\emph{curr}) < \cost(C_\emph{best})$}
          \State $C_\emph{best} \gets C_\emph{curr}$ 
          \State $\error_\emph{best} \gets \error_\emph{curr}$
        \EndIf
      \EndWhile
      \State \Return $C_\emph{best}$
      \EndProcedure
    \end{algorithmic}
\end{algorithm}

\subsection{How to implement \ourscaps efficiently} 
We now discuss key implementation ideas that improve the performance of \ours
in practice.

\paragraph{Weighing fast \& slow} Applying a transformation to a circuit is decomposed into selecting a transformation to apply and a location in the circuit
to apply it to. 
In practice, we limit the probability of choosing resynthesis to 1.5\% of the time, since it is expensive.
In the remaining 98.5\% of the time, we uniformly sample one of the rewrite rules.

\paragraph{Randomly selecting subcircuits}
We choose a random subcircuit to apply the transformation to by picking a node uniformly at random 
in the circuit DAG to begin constructing a subcircuit from. For rewrite rules transformations, 
completely random subcircuits will likely not be transformed nontrivially.
We optimize this step by starting at a random node and performing a full pass through the circuit, replacing every 
disjoint match of the left-hand side with the right-hand side of the rule. 
For resynthesis transformations, 
we start at a random node and grow a subcircuit greedily until we cannot add more nodes 
without exceeding the qubit limit. We only apply resynthesis to a single subcircuit per iteration.

\paragraph{Applying resynthesis asynchronously}
Invoking a unitary synthesis subroutine, even for a circuit with only 3 qubits, 
is slow and takes on the order of seconds or minutes to return a solution. 
To make better use of our time, we choose to make these calls asynchronously
so we can apply rewrite rules concurrently.
If the result of resynthesis is accepted, we effectively discard all modifications from rewrite rules 
made in the interim.

\begin{table}[t]
    \centering
    \small
    \begin{tabular}{lll}
            \toprule
            Gate set & Gates & Architecture \\
            \midrule 
            \ibm  \cite{qiskit}                       & $U1^\theta, U2^{\theta_1, \theta_2}, U3^{\theta_1, \theta_2,\theta_3}, CX$ & Supercond. \\
            \newibm  \cite{qiskit}                    & $\rz{}, SX, X, CX$ & Supercond. \\
            \ion  \cite{ionqGateset}                        & $R_x^\theta, R_y^\theta, \rz{}, R_{xx}^\theta$ & Ion Trap \\
            Nam \cite{nam2018automated}       & $\rz{}, H, X, CX$ & None \\
            \cliffordt \cite{gottesman1998cliffordt} & $T, T^\dagger, S, S^\dagger, H, X, CX$ & Fault Tolerant\\
            \bottomrule
    \end{tabular}
    \caption{Summary of gate sets.}
    \label{tbl:gatesets}
\end{table}

\section{Implementation and Evaluation}
\label{sec:eval}
We implemented \ours in Java.
\ours interfaces with existing resynthesis tools \cite{younis2021bqskit,paradis2024synthetiq} and can be instantiated with arbitrary rewrite rules and gate sets.
We designed our evaluation of \ours to answer the following research questions:
\begin{description}
    \item[Q1] How does \ours compare to state-of-the-art tools?
    \item[Q2] What's the effect of unifying rewriting \& resynthesis?
    \item[Q3] What's the best way to apply rewriting \& resynthesis?
    \item[Q4] Does \ours extend to fault-tolerant computing (\ftqc)?
\end{description}
We focus on \nisq in the first three questions using the diverse gate sets and tools available and then explore \ftqc in Q4. 

\paragraph{Gate sets} Our approach is flexible and can handle arbitrary gate sets. 
We evaluate on a variety of gate sets for promising quantum architectures, summarized in \cref{tbl:gatesets}. 
The Nam gate set is not realized directly on hardware but is studied extensively in prior work due to its resemblance 
to the \cliffordt gate set.

\paragraph{Benchmarks} We consider all the benchmarks used in prior optimization
work \cite{xu2023queso, li2024quarl} as well as benchmarks used in approximate optimization
\cite{patel2021QUEST}. 
Prior optimization work primarily focuses on circuits with fewer than \numprint{2000}
gates. 
We expand the suite to larger application circuits considered in prior mapping-and-routing
work \cite{wille2018jkumapping},
and circuits implementing standard algorithms.
Experimenting with larger circuits is key because total gate count is the primary metric that affects the scalability of optimizers for circuits with more than a few qubits.
Our benchmark suite of 247 circuits includes 
important quantum algorithms in the near and long term such as 
QAOA~\cite{farhi2014qaoa}, VQE~\cite{peruzzo2014vqe}, QPE~\cite{kitaev1995qpe}, 
QFT~\cite{coppersmith2002qft}, Grover's~\cite{grover1996}, and Shor's~\cite{shor1994}. 
To ensure a fair comparison between 
each tool's optimization phase, the input circuit throughout this evaluation
is always already decomposed into the target gate set. \cref{fig:benchmark-sizes} in \cref{sec:benchmark-data}
summarizes the total gate counts of all the input circuits. The benchmark circuits act on 4 to 36 qubits.

\paragraph{Metrics} For \nisq, we focus on two-qubit gate reduction because two-qubit gates have 
orders of magnitude higher error rates compared to single-qubit gates. Gate reduction is 
computed as $1 - \frac{\text{optimized count}}{\text{original count}}$. We also 
compute the circuit \emph{fidelity}, or success probability, to emphasize that 
two-qubit gates are the dominant source of error. The fidelity of a gate
is $1-$ its error rate and the fidelity of a circuit is the product of its gate fidelities.
For $\ibm$ and $\newibm$, 
we use the calibration data for the IBM Washington device available in Qiskit;
for the Ion gate set, we use data for the IonQ Forte device \cite{forteIonQ}.
In Q4, we consider different metrics for \ftqc. 

\emph{In our plots, each point corresponds to a benchmark. 
For each benchmark circuit and tool, we compute the mean metric for a number of runs of the tool (10 trials for \ours)
and a 95\% confidence interval.}
For readability, we present the benchmarks in increasing order sorted based on \ours.
A point where \ours lies above the respective point for the other tool implies 
that \ours outperforms the other tool. The bar plot below each scatter plot summarizes 
the number of benchmarks \ours on average outperforms, matches, 
and underperforms, respectively, the tool in the title.

\subsection*{Q1: Comparison with state-of-the-art optimizers}

\begin{figure*}[t!]
    \begin{subfigure}{0.95\linewidth}
        \centering
        \includegraphics[width=\textwidth]{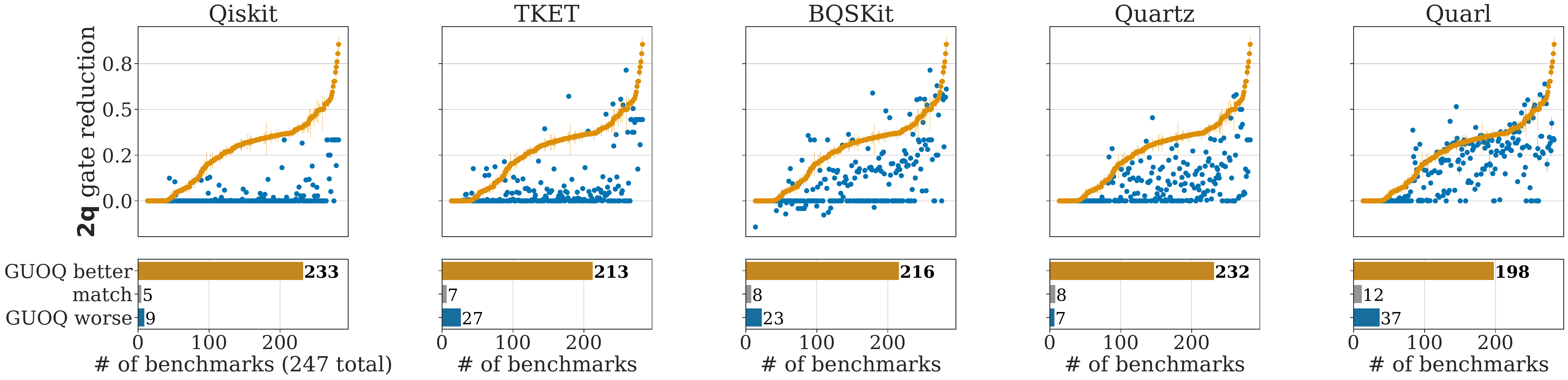}
        \includegraphics[width=\textwidth]{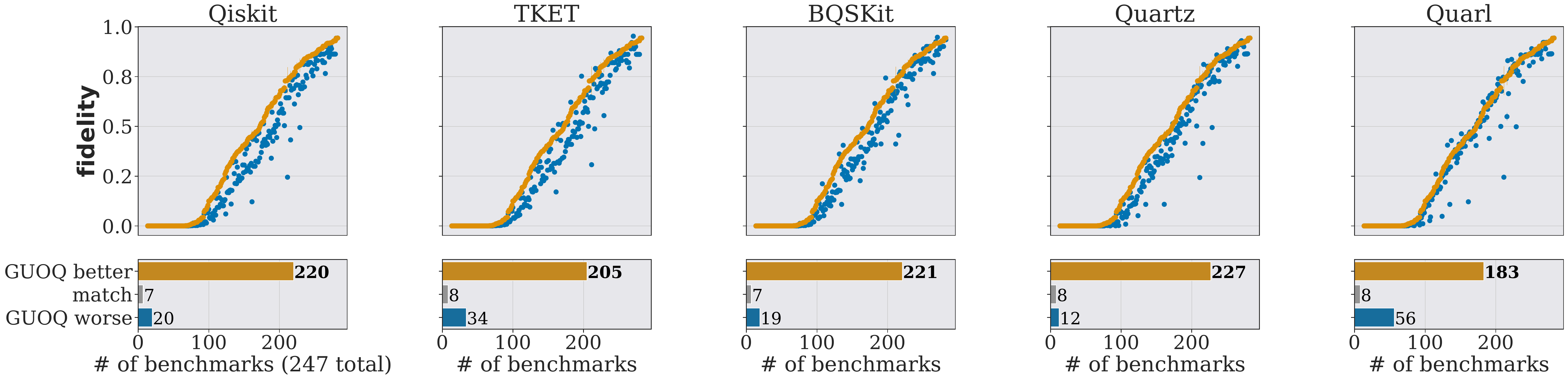}

    \end{subfigure}
    \caption
    {Comparison against state-of-the-art optimizers on the $\newibm$ gate set. Each graph has a 
    bar chart summarizing the number of benchmarks \ours outperforms, matches, or underperforms the tool in the title.
    In each graph, orange is \ours and blue is the tool in the title. 
    Each point is the mean metric and a 95\% confidence interval for a number of runs on a single benchmark.
    The points are sorted based on \ours and a point where the orange lies above the blue indicates \ours is better. } 
    \label{fig:rq1_ibmn_super}
\end{figure*}

\begin{table}[t]
    \centering
    \smaller
    \begin{tabular}{lcp{4cm}}
            \toprule
            Tool & Superoptimize & Approach \\
            \midrule
            Qiskit \cite{qiskit}            & \xmark & fixed sequence of passes\\
            \tket \cite{tket}               & \xmark & fixed sequence of passes\\
            \voqc \cite{hietala2021voqc}    & \xmark & fixed sequence of passes\\
            \bqskit \cite{younis2021bqskit} & \cmark & partition + resynthesize\\
            \queso \cite{xu2023queso}       & \cmark & beam search + rewrite rules\\
            Quartz \cite{xu2022quartz}      & \cmark & beam search + rewrite rules\\
            Quarl \cite{li2024quarl}        & \cmark & reinf. learning + rewrite rules\\
            \bottomrule
    \end{tabular}
    \caption{State-of-the-art optimizers.}
    \label{tbl:rq1-tools}
\end{table}

\begin{figure}[h]
    \begin{subfigure}{\linewidth}

        \includegraphics[width=\textwidth]{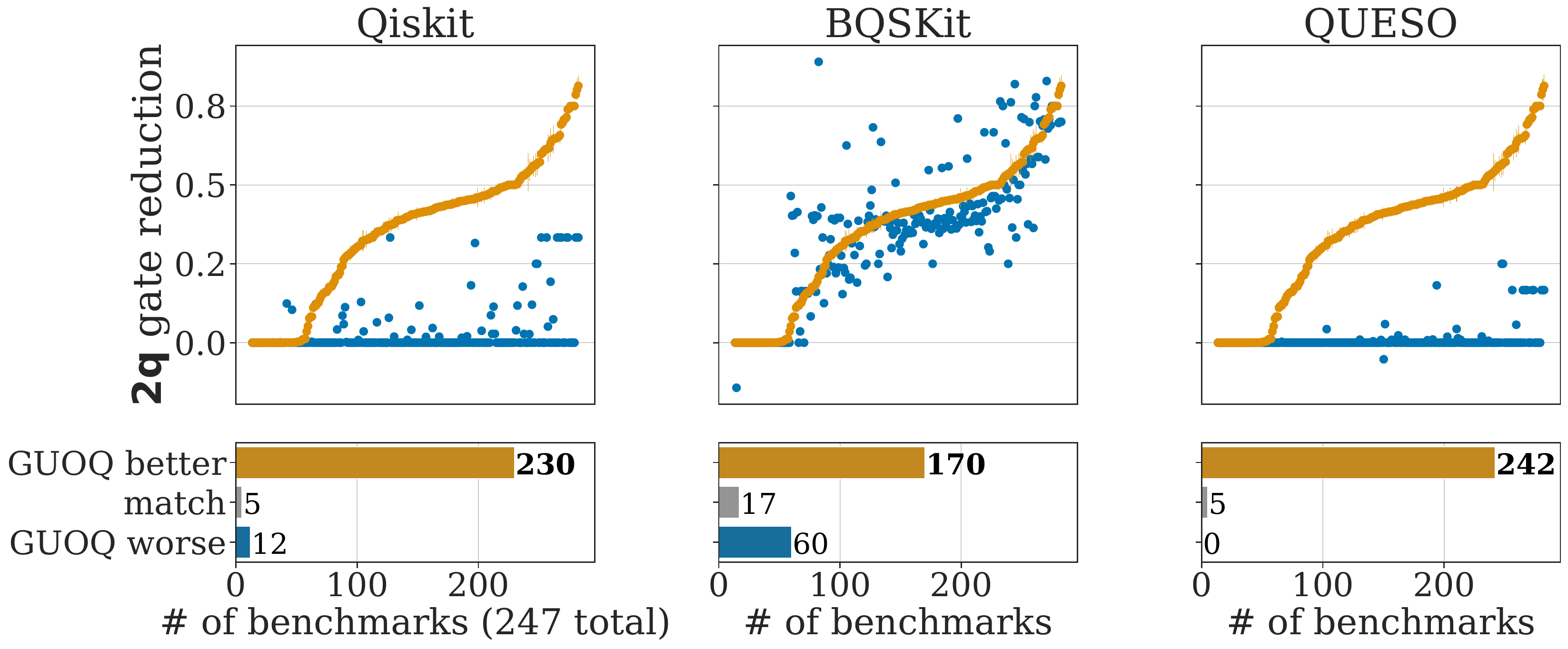}
        \includegraphics[width=\textwidth]{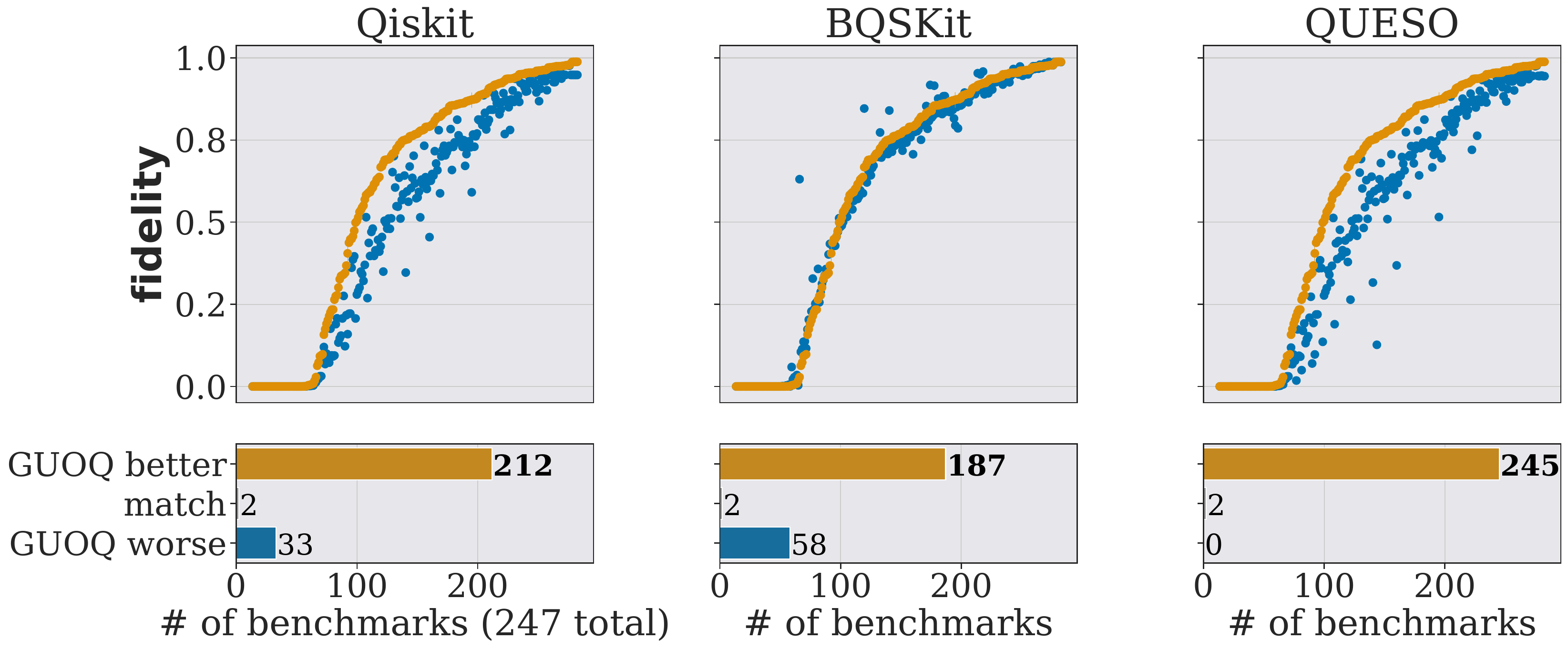}

    \end{subfigure}
    \caption
    {\ours vs state-of-the-art on the $\ion$ gate set.} 
    \label{fig:rq1_ion}
\end{figure}

\paragraph{State of the art} We compared \ours to 7 state-of-the-art optimizers listed in \cref{tbl:rq1-tools}.
Our goal is to compare the optimization phase of various tools so we do not invoke
mapping and routing and tools are allowed to change the input circuit's connectivity
if they support this feature. 
We exclude \pyzx~\cite{kissinger2020Pyzx} from this research question because its primary optimization
objective is to reduce $\tgate$ gate count and often either increases or preserves the 
two-qubit gate count. We compare against \pyzx in Q4 where we explore $\tgate$ 
reduction.

\paragraph{Instantiation of \ours} \ours uses rewrite 
rules generated by \queso~\cite{xu2023queso} and does not consider any
size-increasing rewrite rules. 
To reduce two-qubit gate count, \ours uses \bqskit~\cite{younis2021bqskit} 
for resynthesis and the optimization objective is to maximize fidelity.
We limit random subcircuits to have at most 3 qubits.
The temperature hyperparameter $t$ is set to $10$, corresponding to a very small probability of accepting a worse solution.
We chose this value empirically by performing a sweep of values from 0 (always accept) to 10. 

\paragraph{Experimental setup} 
Unless otherwise indicated, we allocated each tool 1 hour, 32GB of RAM, and 1 
CPU core on an AMD EPYC 7763 64-Core Processor.
Quarl was run on a cluster of machines and was allocated 64GB of RAM, 1 NVIDIA A100-SXM4-40GB or 80GB GPU, and 1 CPU core.
We only evaluate a tool on its supported gate sets. 
For approximate tools, we 
enforce an error upper bound of $10^{-8}$, which is (1) many orders of magnitude smaller
than the error rate of a single two-qubit gate in \nisq ($10^{-3}$~\cite{khan2024quantinuumfidelity}) and (2) on-par with
the logical error rate of a single error correction cycle for \ftqc($10^{-6}$ to $10^{-9}$~\cite{neven2023googleft}) or an arbitrary-angle approximation. 

We run Quarl for 3 trials with rotation merging, following 
their paper's experimental setup. For Qiskit and \bqskit, we 
use the most powerful optimization levels, which are 3 and 4, respectively.
We report the best solution found within the time and memory limits for all tools. 
That is, we use the partial solutions that \ours, \queso, Quartz, and Quarl provide 
and use the original circuit for the other tools.

\paragraph{Results} The results for the $\ibm$, $\newibm$, and Nam gate sets are 
all similar, so we only show the plots for the $\newibm$ gate set. 
\cref{fig:rq1_ibmn_super} shows the comparison between \ours and state-of-the-art with respect to two-qubit gate reduction
and fidelity on the \newibm gate set. Recall that benchmarks where the orange point lies above 
the blue point are ones where \ours outperforms the tool in the title. 
For example, consider the left-most column where
\ours outperforms Qiskit with respect to two-qubit gate reduction on 233/247 of 
the benchmarks, matches on 5, and underperforms on 
9. We observe a very similar story in the fidelity graph on the bottom. 
This result holds in general against all other tools and \ours 
outperforms state-of-the-art on \emph{at least} 80\% and 74\% of the benchmarks with respect to two-qubit gate reduction
and fidelity, respectively. Recall that Quarl requires a GPU to run its specialized reinforcement learning and trains on a subset of the benchmark suite. 
\ours reduces two-qubit gate count by 28\% on average while the next best tool, Quarl, has an average reduction of 18\% and 
the best industrial toolkit, \tket, achieves 7\% average reduction. 

The results in \cref{fig:rq1_ion} on the \ion gate set depict a similar story. 
As we can see in the comparison against \queso,
a tool which synthesizes rewrite rules,
the \ion gate set is challenging for \queso because rewrite rules are limited to patterns 
with a maximum of 3 gates to limit the combinatorial explosion of rules. 
\ours performs well because resynthesis can compensate for the limited rewrite rules.
We will see in Q4 an example of the opposite effect. This demonstrates how the effectiveness of rewrite rules and resynthesis 
varies across different gate sets. Thus, unifying them provides a generic approach for optimizing diverse circuits.

\begin{mybox}
    \paragraph{\emph{Q1 summary}} \textbf{\ours significantly outperforms state-of-the-art across all
    gate sets. In the \emph{worst} case, \ours outperforms other tools on 69\% and 74\% of the benchmarks 
    with respect to two-qubit gate reduction and fidelity, respectively. In particular, \ours achieves an average of 28\% 
    two-qubit gate reduction on the \newibm gate set while the state-of-the-art superoptimizer 
    (requires GPU) and industrial toolkit achieve average reductions of 18\% and 7\%, respectively.}
\end{mybox}

\subsection*{Q2: Effect of combining rewriting  and resynthesis}
\begin{figure}[t]
    \centering
    \includegraphics[width=0.4\textwidth]{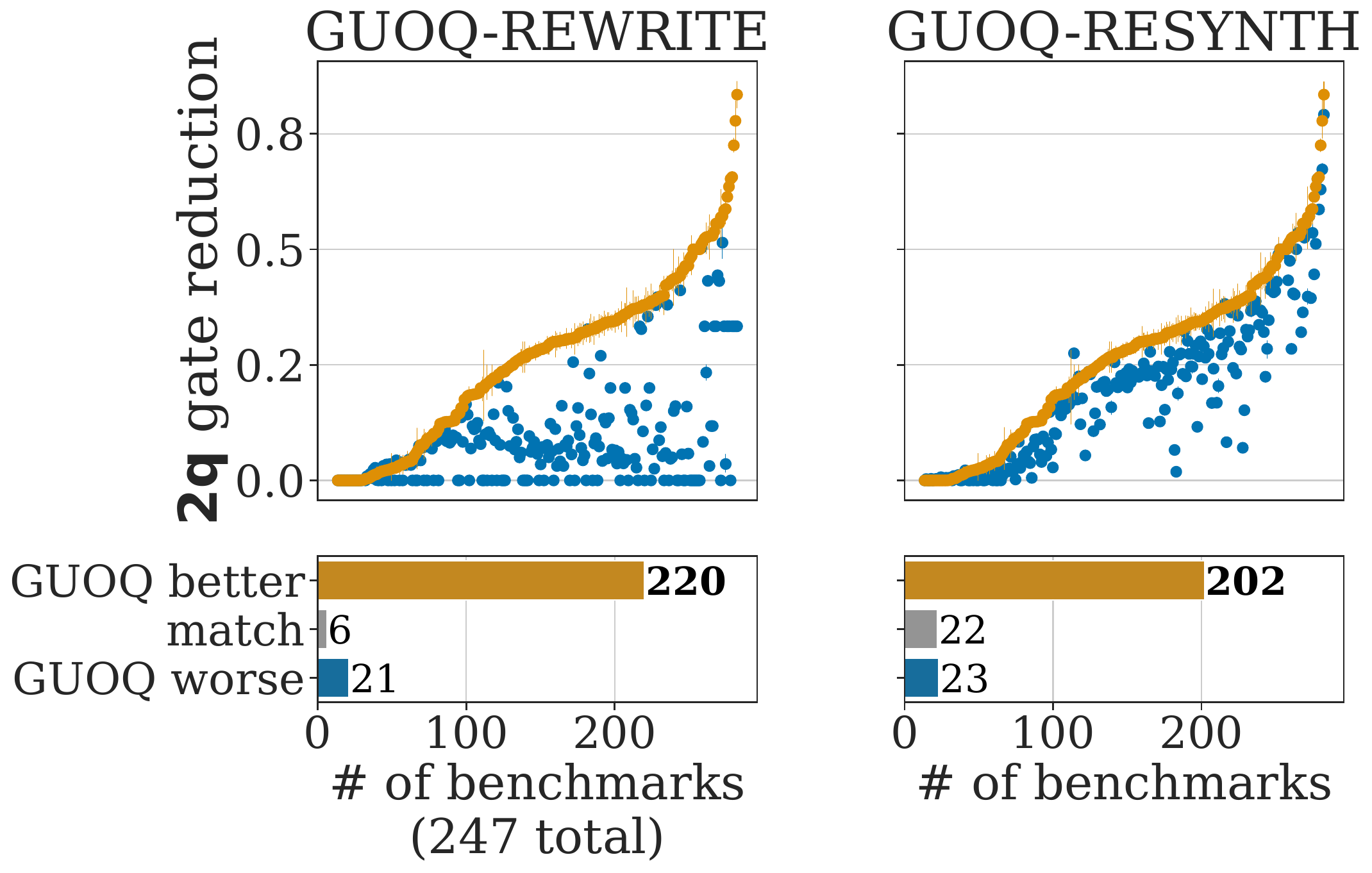}
    \caption
    { \ours vs using only rewrite rules or resynthesis. } 
    \label{fig:rq2}
  \end{figure}

We explored this question by fixing the \ibm gate set and running \ours with 
three different sets of transformations.
\oursrr only uses rewrite rules synthesized by \queso, \oursresynth only uses 
resynthesis, and \ours uses both.

\paragraph{Results} \cref{fig:rq2} shows that removing either rewrite rules 
or resynthesis is overall detrimental to the performance of \ours. Similar to 
\cref{fig:rq1_ibmn_super}, the title of each plot is the baseline and points below the 
curve are benchmarks where \ours outperforms the baseline. 
Observe how most of reduction comes from using resynthesis because it is a powerful optimization 
on its own. Interleaving with rewrite rules, which can take care of simple optimizations 
quickly, pushes the reduction even further.

\begin{mybox}

\paragraph{\emph{Q2 summary}} \textbf{We can exploit the synergy between 
rewrite rules and resynthesis to achieve well beyond the capabilities of either alone.}
\end{mybox}
\subsection*{Q3: How to combine rewriting and resynthesis?}
\label{sec:q3}
\begin{figure}[t]
    \centering
    \begin{subfigure}{\linewidth}
        \includegraphics[width=\textwidth]{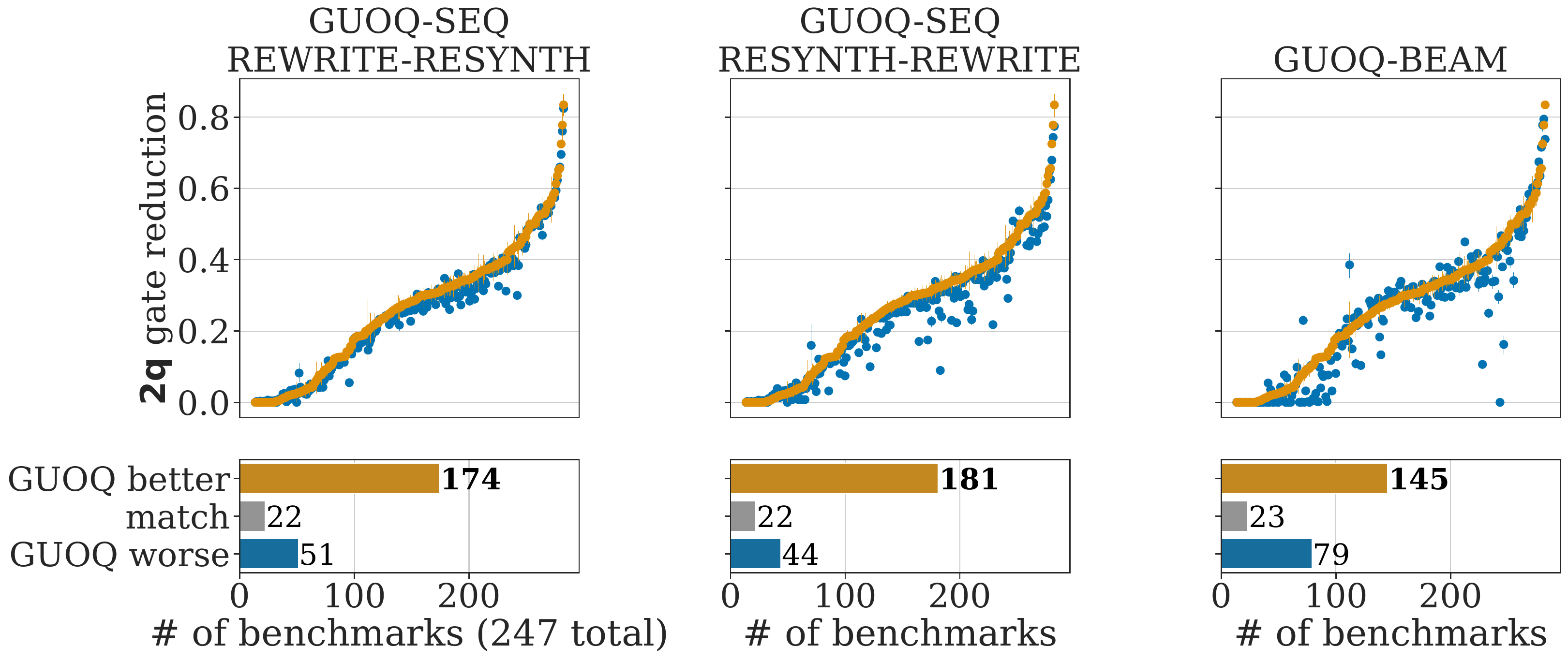}
    \end{subfigure}
    \caption
    { Comparing \ours against other search algorithms} 
    \label{fig:rq3}
  \end{figure}

  \begin{figure*}[t!]
    \begin{subfigure}{0.95\linewidth}
    \centering
    \includegraphics[width=\textwidth]{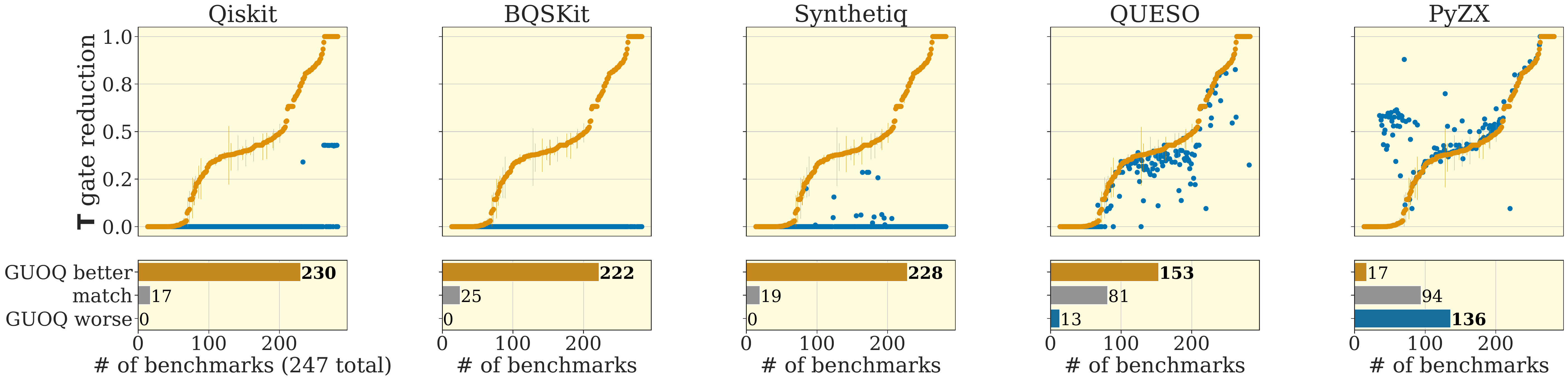}
    \includegraphics[width=\textwidth]{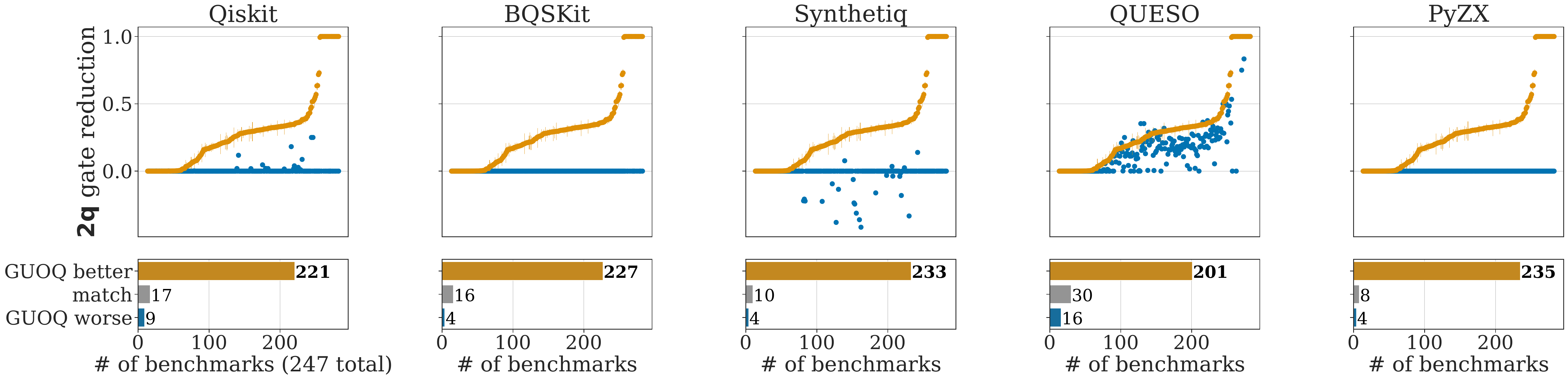}
    \end{subfigure}
    \caption{ Comparison against state-of-the-art optimizers on the \cliffordt gate set.} 
    \label{fig:rq4}
\end{figure*}

To answer this question, we fixed the \ibm gate set and set of rewrite rule and resynthesis transformations
while varying the search algorithm. 
We compare \ours against three alternate search algorithms 
for combining rewrite rules and resynthesis:
(1) \oursrrfirst,
which spends the first half of the allotted time running \ours with rewrite rules only, then switches to running with resynthesis only,
(2) \oursresynthfirst does the opposite,
and (3) \textsc{guoq-beam} uses the MaxBeam algorithm of \queso~\cite{xu2023queso} to instantiate our framework.

\paragraph{Results} 
\cref{fig:rq3} shows the results.
\ours outperforms the coarse 
interleaving \oursresynthfirst and \oursrrfirst use on a majority of the benchmarks. This implies that 
tightly interleaving these different transformations is preferable to choosing a fixed ordering.
We  also see that relative to each other: \oursresynthfirst and \oursrrfirst 
result in different solutions, which is further evidence 
that the ordering matters. 

We now turn our attention to \oursbeam, which allows arbitrary orderings of rewrite rules and 
resynthesis but does not randomly sample transformations like \ours. Instead, \oursbeam maintains 
a large bounded priority queue of candidates and attempts to apply every transformation in each iteration.
Doing so generates many candidates in each iteration (one for each transformation successfully applied),
which saturates the queue quickly with solutions of the same cost. In fact, the solutions 
are generally a few local transformations away from one another so the search makes much slower progress compared to \ours.
The influx of candidates to the bounded queue also causes many solutions to be pruned, effectively wasting 
the time spent generating those candidates. 
Especially for larger circuits, the sizable queue is memory intensive, causing further 
slowdowns. 
In summary, the benefit of beam search considering many candidates to avoid 
local minima is lost in this problem setting.

\begin{mybox}

    \paragraph{\emph{Q3 summary}} \textbf{\ours achieves the best results by tightly interleaving rewrite rules and resynthesis using a simple, lightweight randomized algorithm.  }

\end{mybox}

\subsection*{Q4: Does \ours extend to \ftqc?}
\label{sec:q4ft}

In this research question, we shift our focus
to the fault-tolerant \cliffordt gate set, where the desired optimization 
objective is an amalgamation of two NP-hard optimization problems~\cite{vandewetering2024optimisingquantumcircuitsgenerally}. 
We want to primarily reduce $\tgate$ gates~\cite{campbell2017roadtoft} and reducing two-qubit
gates is secondary, but still critical. Error correction is not perfect and 
the longer a quantum computation runs, the 
higher the risk of accruing uncorrectable logical error.
Two-qubit gates, specifically $\cx$ in the \cliffordt gate set,
increase the circuit runtime disproportionately because they inherently require
more time compared to single-qubit Clifford gates \cite{litinski2018latticesurgery} and architectural 
constraints can limit parallel execution \cite{hua2021autobraid,beverland2022edpc}.
Furthermore, we anticipate that the problem of $\cx$ congestion will be exacerbated in 
compact \ftqc architectures \cite{litinski2018gameofsurface} with less routing space.
Lastly, recent work~\cite{gidney2024magicstatecultivationgrowing} has significantly reduced the space-time cost of 
preparing a $\tgate$ magic state, indicating that the focus on solely reducing $\tgate$ gates 
may soon need to be recalibrated.

We instantiate \ours with \synthetiq~\cite{paradis2024synthetiq}, a 
state-of-the-art unitary synthesis algorithm for finite gate sets, and rewrite rules generated by \queso~\cite{xu2023queso}.
We consider two additional tools in this comparison: (1) \pyzx~\cite{kissinger2020Pyzx}, a state-of-the-art optimizer for 
reducing $\tgate$ count using the rewrite-rule-based ZX-calculus and (2) our implementation of a \bqskit-style partitioning optimizer~\cite{patel2021QUEST}
that uses \synthetiq. 

\begin{figure}[t]
    \centering
    \includegraphics[width=0.4\textwidth]{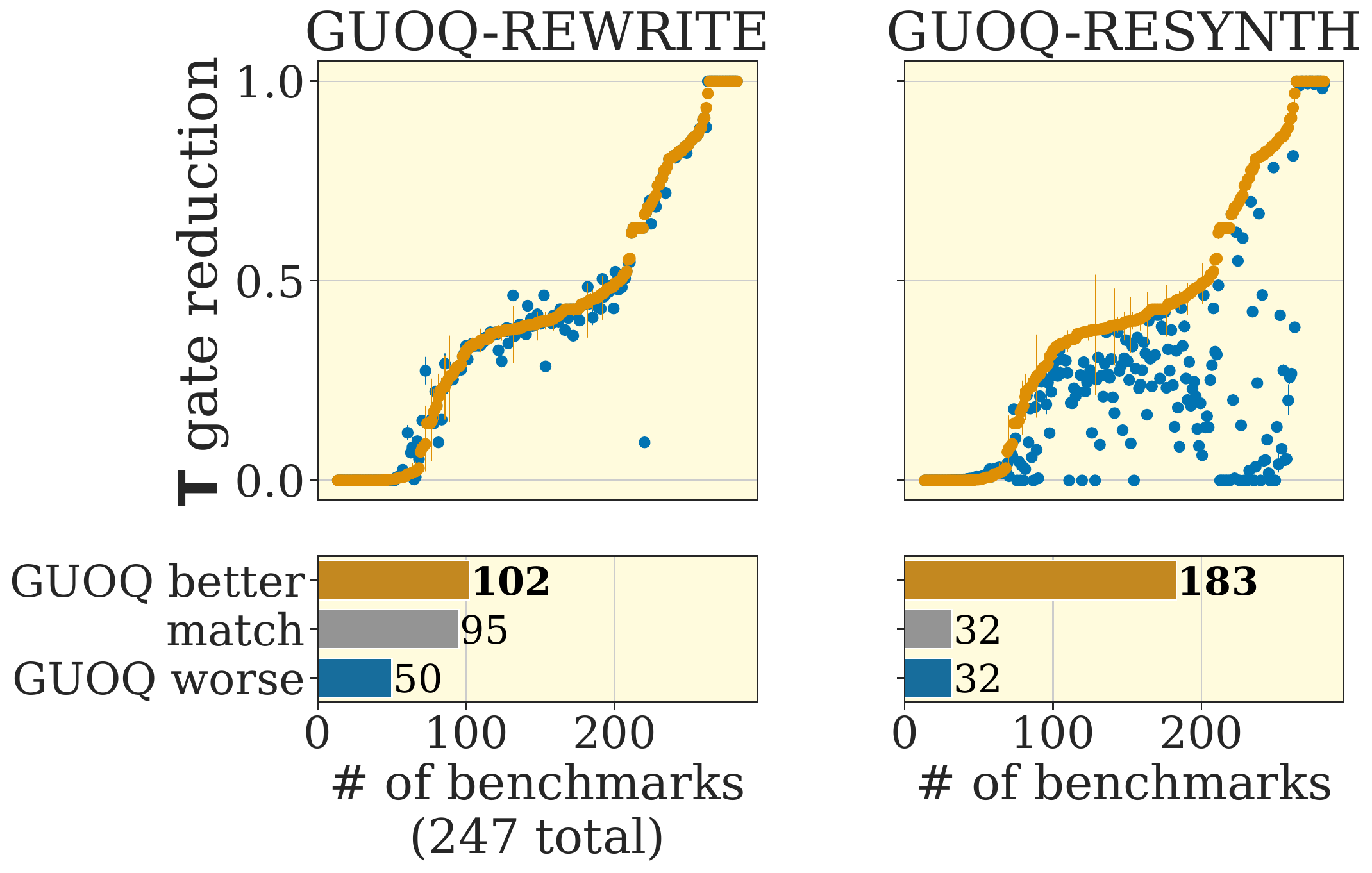}
    \caption
    { Revisiting Q2 for the \cliffordt gate set. } 
    \label{fig:rq4-rq2}
\end{figure}

\begin{figure}[t]
    \begin{subfigure}{0.47\linewidth}
        \centering
        \includegraphics[width=\textwidth]{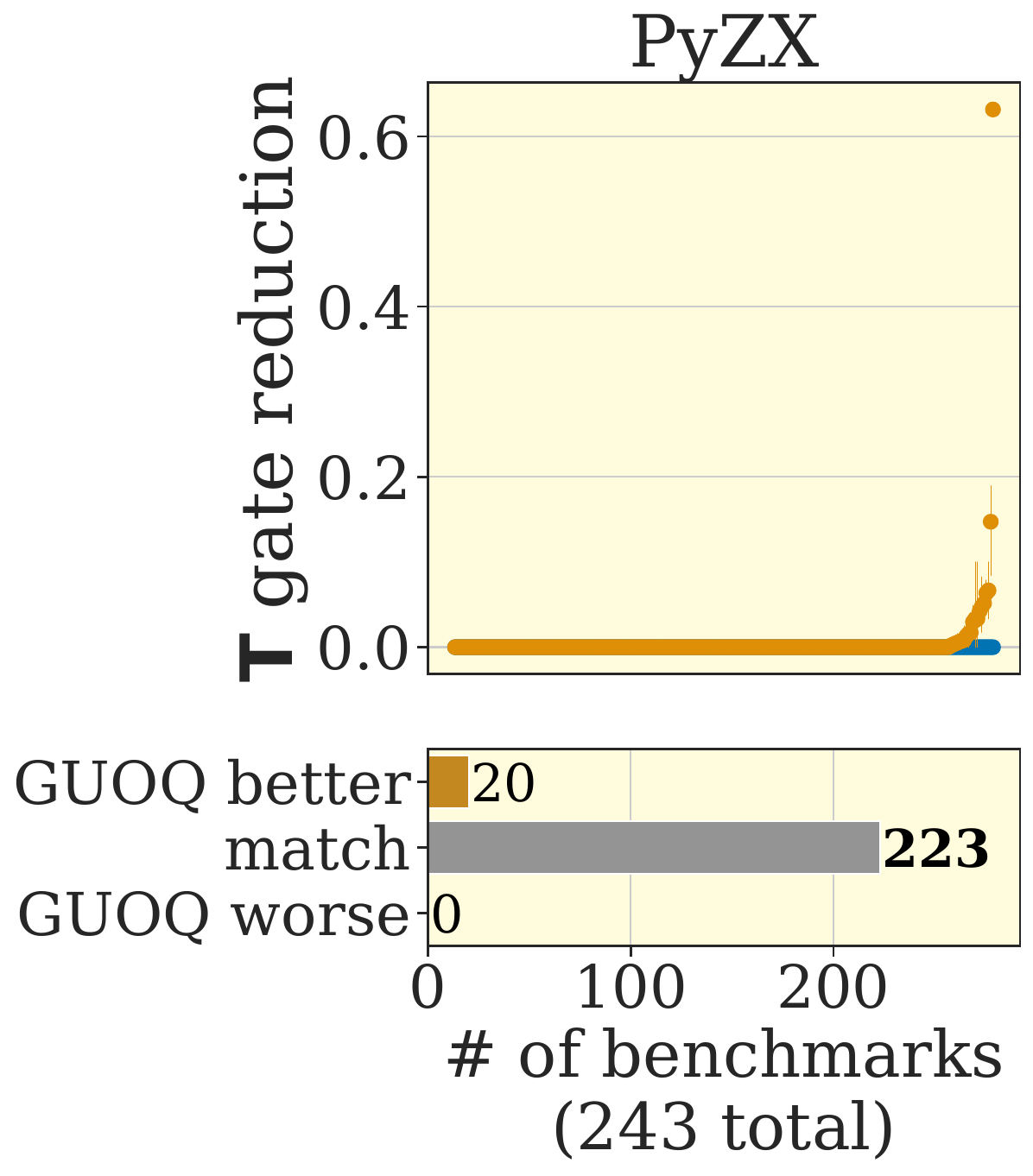}
    \end{subfigure}
    \begin{subfigure}{0.47\linewidth}
        \centering
        \includegraphics[width=\textwidth]{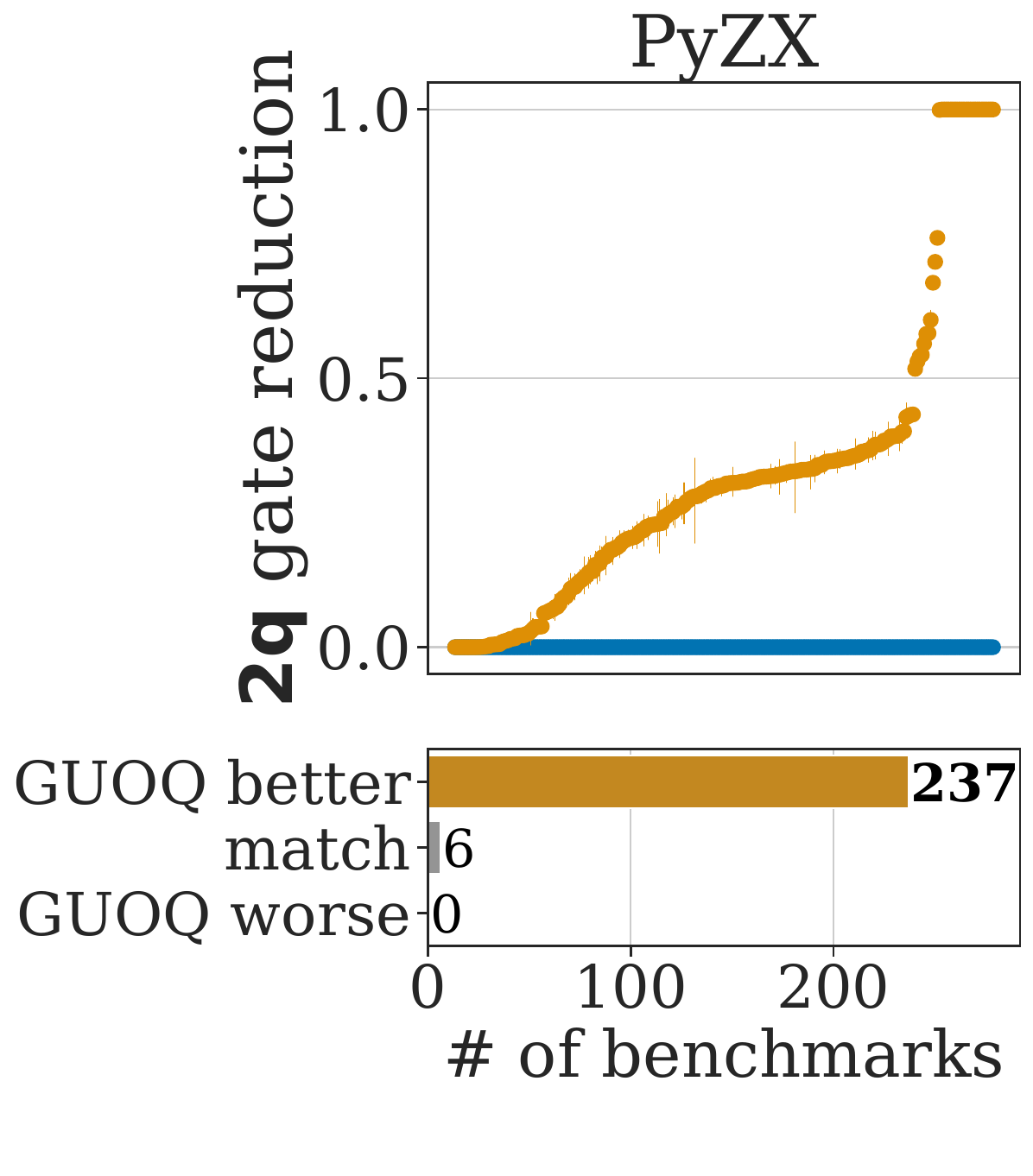}
    \end{subfigure}
    \caption{ Running \ours on \pyzx output. } 
    \label{fig:rq4_after_pyzx}
\end{figure}

\paragraph{Results} The top and bottom rows of \cref{fig:rq4} show the comparison against 
other tools with respect to $\tgate$ and $\cx$ gate reduction respectively.
\ours outperforms all tools except \pyzx with respect to $\tgate$ reduction and outperforms 
all tools with respect to $\cx$ gate reduction.
Observe how reducing $\tgate$ gates is hard and a general-purpose tool, like Qiskit, only reduces
the number of $\tgate$ gates on 5\% of the benchmarks. 
\pyzx is a domain-specific optimizer that uses a powerful graph-based theory, the ZX-calculus, which 
is specialized for reducing phase gates, but does not reduce $\cx$ gates at all. 

Digging deeper, we find that \ours is unable to surpass \pyzx in $\tgate$ gate reduction 
because unitary synthesis for finite gate sets is much harder than for continuous gate sets. 
\cref{fig:rq4-rq2} shows the results from the same ablation as in Q2
and we find that rewrite rules contribute more than resynthesis.

To get a sense for how ``good'' \pyzx's solution is, we ran \ours on the 
output of \pyzx for the 243 benchmarks it provided a solution for within the time and memory limits. 
We discovered that \ours can drastically reduce the $\cx$ gate count
without increasing $\tgate$ gate count! \cref{fig:rq4_after_pyzx} shows this result,
which is exciting because \pyzx on its own does not reduce $\cx$ gate count. Extending
\pyzx with \ours pushes the boundaries for the multifaceted \ftqc optimization objective.

\begin{mybox}
    \paragraph{\emph{Q4 summary}} \textbf{\ours outperforms all tools with respect to 
    $\tgate$ gate reduction except \pyzx, which it only outperforms or matches on 45\% of the benchmarks.
    However, \ours vastly outperforms \pyzx (and other tools) with respect to $\cx$ gate reduction
    on \emph{at least} 81\% of the benchmarks, which 
    is also critical in \ftqc. Additionally, when run on the output of \pyzx, \ours can 
    reduce $\cx$ gate count on average by 32\% without increasing $\tgate$ gate count.}
\end{mybox}

\section{Related Work}

\paragraph{Quantum optimizers}
Traditional quantum-circuit optimizers primarily use a fixed set of 
hand-crafted optimizations \cite{nam2018automated,hietala2021voqc, tket, qiskit, amy2020staq, campbell2023superstaq, paykin2023pcoast}
applied in a fixed sequence. \pyzx \cite{kissinger2020Pyzx} takes a variant of this approach: representing a circuit as a ZX-diagram and applying the 
graphical rewrite rules of the ZX-calculus. \ours instead performs a fine-grained search over arbitrary optimizations.

\paragraph{(Quantum) superoptimizers}
The idea of \emph{superoptimizing} classical programs has been around for decades~\cite{turchin1986supercompiler,massalin1987super}. 
This vast line of work~\cite{schkufza2013stoke,phothilimthana2016scaling,churchill2017loopsuper,sasnauskas2017souper,mukherjee2020soupertwo,liu2023minotaur,jia2019taso} 
stems from the idea of finding the \emph{optimal} solution for small programs that can later be applied in peephole optimizers.
Approaches like STOKE~\cite{schkufza2013stoke}
use MCMC~\cite{hastings1970mcmc} to superoptimize x86 assembly by randomly mutating the program. 
The ergodic theory behind MCMC lends itself to applications where testing correctness is easy.
However, quantum circuits cannot be
efficiently simulated on classical hardware \cite{nielsen2002quantum}.

Rewrite rules and unitary synthesis have both been used as the basis of quantum-circuit \emph{superoptimizers}.
Quartz \cite{xu2022quartz} and \queso \cite{xu2023queso} synthesize rewrite rules for a given gate set and use beam search to explore 
the space of rule application schedules. Quarl~\cite{li2024quarl} applies reinforcement learning to 
schedule the application of rules generated by Quartz. None of these apply approximate circuit transformations. 
On the other hand, resynthesis-based superoptimizers \cite{patel2021QUEST,younis2021bqskit}
optimize circuits by performing a single pass of partitioning into subcircuits, followed by applying unitary synthesis to each subcircuit.
This approach circumvents the qubit count limitations of unitary synthesis, but is rigid and misses
potential optimization opportunities that straddle the boundary between two adjacent partitions.
In contrast, \ours is not limited to resynthesizing disjoint subcircuits of the original circuit. \ours can freely choose subcircuits to resynthesize by using \cref{theorem:error_bound}
to bound the error when composing applications of resynthesis. 

In a similar spirit to Quarl, other recent approaches have applied reinforcement learning to superoptimize quantum circuits.
MQTPredictor~\cite{quetschlich2024mqtpredictor} predicts the optimal passes and device 
with respect to an optimization objective but currently only considers a subset of Qiskit and \tket passes.
AlphaTensor-Quantum~\cite{ruiz2024alphatensor} is a closed-source approach that 
uses reinforcement learning for tensor decomposition to optimize $\tgate$ count.

\paragraph{Domain-specific optimizers}
Other work targets specific applications like Hamiltonian simulation \cite{li2022paulihedral, liu2024fermihedral}
or variational algorithms \cite{jin2023tetris}.
Some tools operate at a lower level of abstraction than we do by 
considering gate \emph{pulses} \cite{shi2019aggregate} or a higher level by starting from a
program written in a high-level quantum programming language~\cite{yuan2022tower, yuan2023spire}.
In contrast, \ours is designed to be flexible for diverse quantum 
assembly circuits and architectures.

\paragraph{Unitary synthesis}
Extensive prior work has considered the unitary synthesis 
problem for both finite and parameterized
gate sets. For finite gate sets, some approaches~\cite{tucci2005kak,amy2014middle}
provide theoretical guarantees of optimality in terms of circuit size, whereas others
\cite{kang2023modular,paradis2024synthetiq} sacrifice optimality for improved runtime. 
For parameterized gate sets, 
several techniques~\cite{davis2020qsearch,smith2023leap,younis2021qfast,rakyta2022squander,rakyta2022squandertwo} 
use numerical optimization to instantiate template circuits. However, all of these algorithms can only be applied to circuits
with a handful of qubits.

\section{Conclusions}
We have described a generic and flexible framework for unifying rewriting and 
resynthesis for quantum-circuit optimization along with a simple and effective algorithm 
parameterized on an instantiation of this framework. Our approach, \ours, outperforms 
state-of-the-art optimizers in both near (\nisq) and long (\ftqc) term quantum computing paradigms.
For future work, we are interested in developing \emph{symbolic} unitary synthesis
so we can learn general transformations on the fly---as opposed to ones with highly specific angles---that
will be more likely to apply later in the search.

\appendix
\section{Proofs}

\paragraph{\cref{theorem:error_bound}}
    By induction on $n$. 
    \emph{Base case:} Trivially, $\circuit_0 \equiv_0 \circuit_0$.
    \emph{Induction case:} Assume $\circuit_0 \equiv_{k \epsilon} \circuit_k$ for $k\geq 0$ as our inductive hypothesis.
        We will show $\circuit_0 \equiv_{(k+1) \epsilon} \circuit_{k+1}$.
        We have $\circuit_k \approxx{} \circuit_{k+1}$ by the proof of \cite[\S3.8]{patel2021QUEST} for disjoint partitions. 
        Let $U \coloneq U_{C_0}$,
        $U' \coloneq U_{C_k}$, $U'' \coloneq U_{C_{k+1}}$, $\epsilon_1 \coloneq k\epsilon$, and $\epsilon_2 \coloneq \epsilon$.
        Now it suffices to show 
        $\hs(U, U'') \leq \epsilon_1 + \epsilon_2 = (k+1) \epsilon$. 
        \begin{footnotesize}
            \begin{align*}
                & \hs(U, U'') \\
                &= \sqrt{1-\frac{\Vert Tr(U^\dag U'')\Vert ^2}{N^2}} && \text{\cref{def:hs}}\\
                &= \sqrt{1-\frac{\Vert Tr(U^\dag U' U'^\dag U'')\Vert ^2}{N^2}} && \text{$U'U'^\dag = I$} \\
                &= \sqrt{1-\frac{\Vert Tr[(U^\dag U')(U'^\dag U'')]\Vert ^2}{N^2}} && \text{Grouping terms} \\
                &\leq \sqrt{1-\frac{\Vert Tr(U^\dag U')\Vert ^2}{N^2}} + \sqrt{1-\frac{\Vert Tr(U'^\dag U'')\Vert ^2}{N^2}} && \text{\cite[\S3.8]{patel2021QUEST}} \\
                &\leq \epsilon_1 + \epsilon_2 && \text{\cref{def:hs,def:approx-equiv}}
            \end{align*}
        \end{footnotesize}

        By definition of
         $\approxx{}$, $\circuit_0 \equiv_{(k+1) \epsilon} \circuit_{k+1}$, as desired. \qedsymbol \renewcommand{\qedsymbol}{}
    
    \noindent
    \paragraph{\cref{theorem:correct}} Follows directly from \cref{theorem:error_bound} and \cref{fig:algs}, line 6.

\section{Benchmark Data}
\label{sec:benchmark-data}
\begin{figure}[!h]
    \centering
    \begin{subfigure}{\linewidth}
        \includegraphics[width=0.49\textwidth]{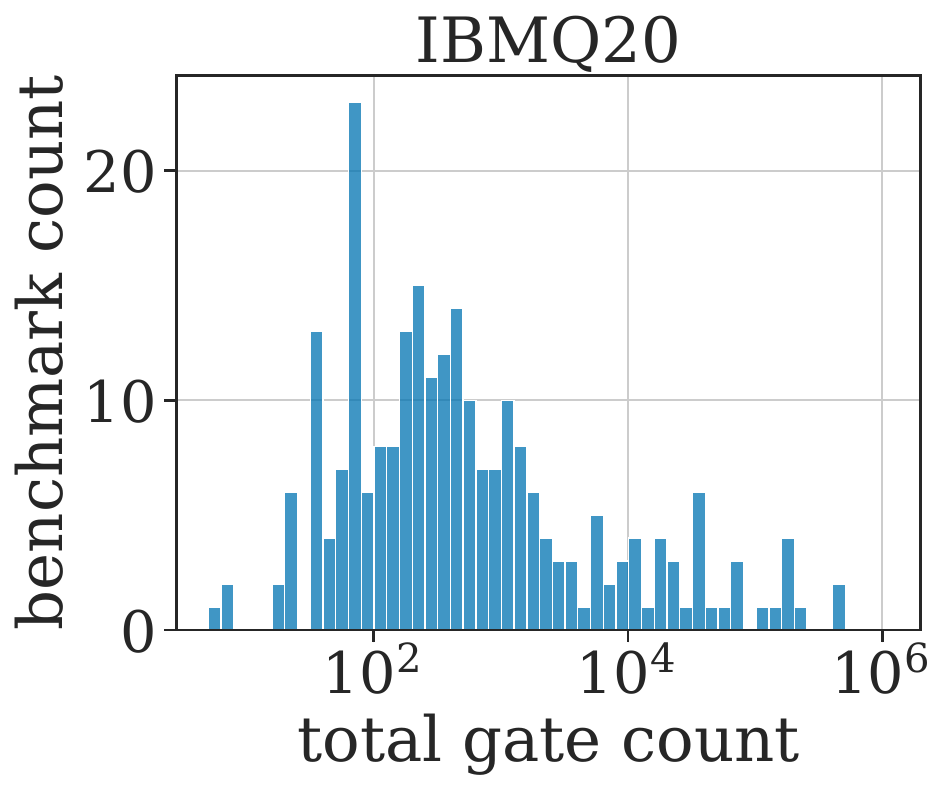}
        \hfill
        \includegraphics[width=0.49\textwidth]{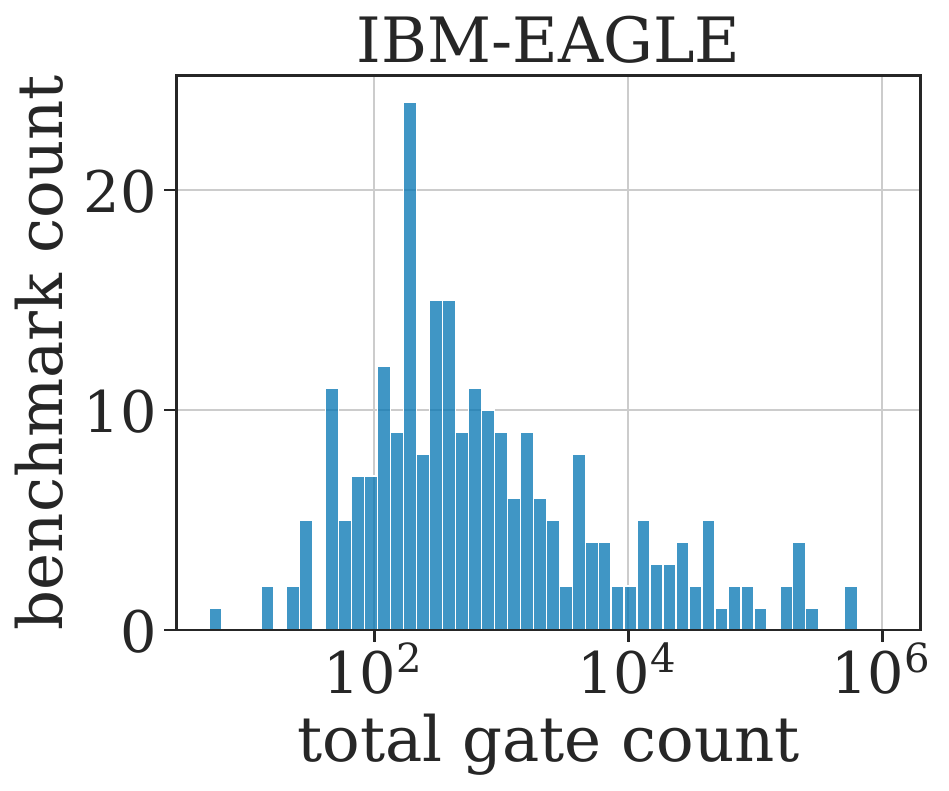}
    \end{subfigure}

    \begin{subfigure}{\linewidth}
    \centering
    \includegraphics[width=0.49\textwidth]{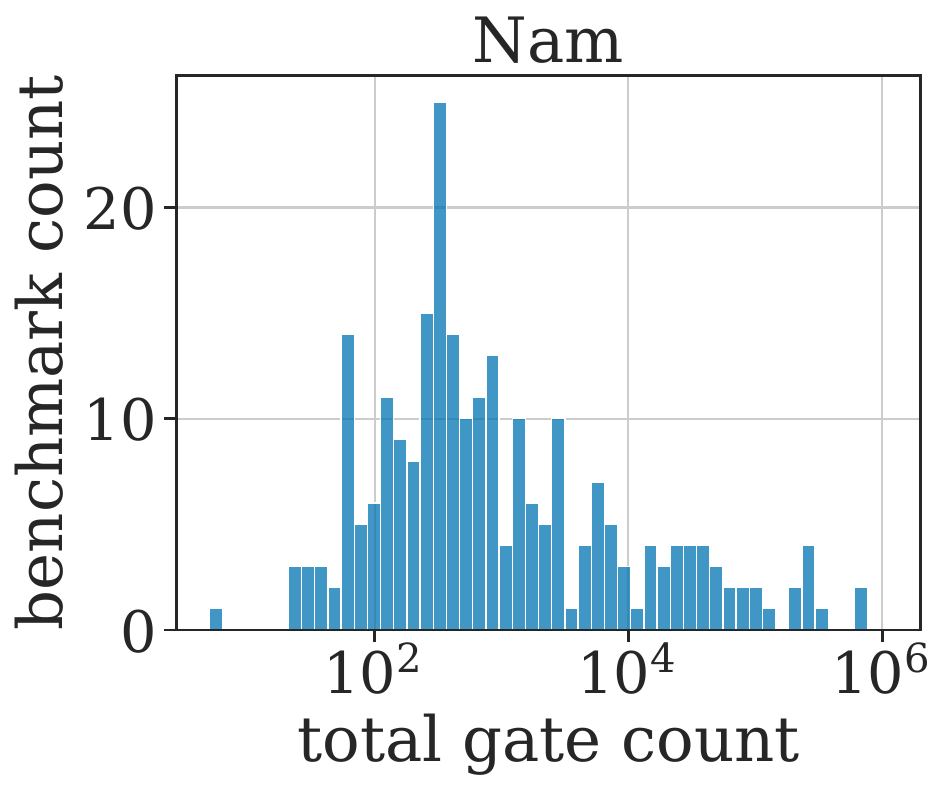}
    \includegraphics[width=0.49\textwidth]{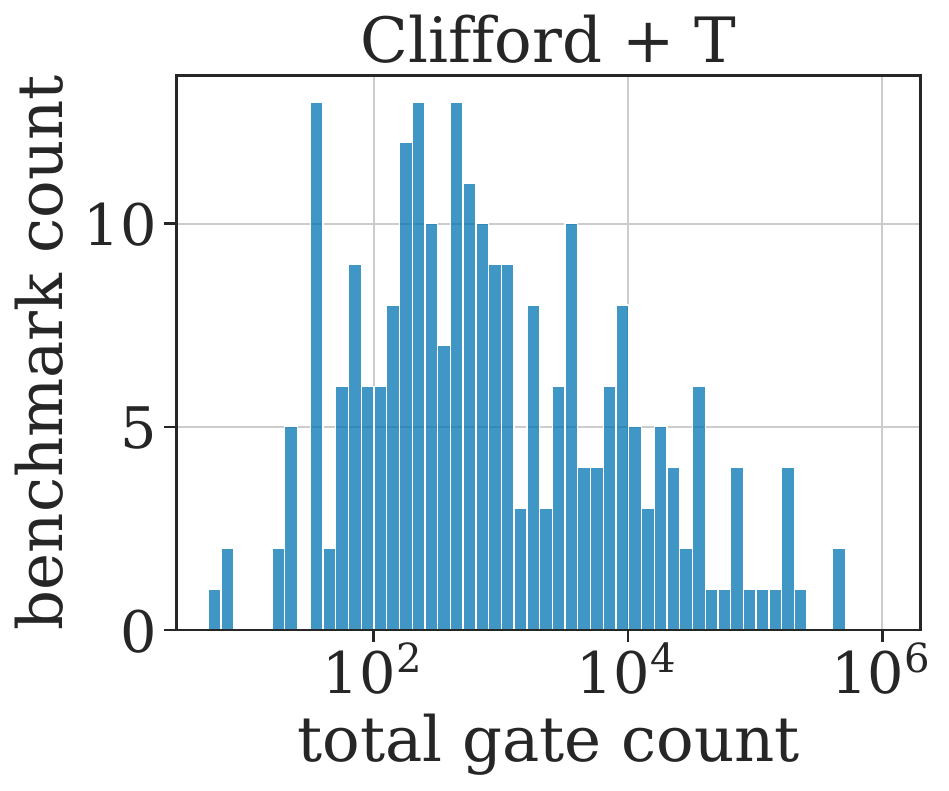}
    \end{subfigure}

    \begin{subfigure}{\linewidth}
        \centering
        \includegraphics[width=0.49\textwidth]{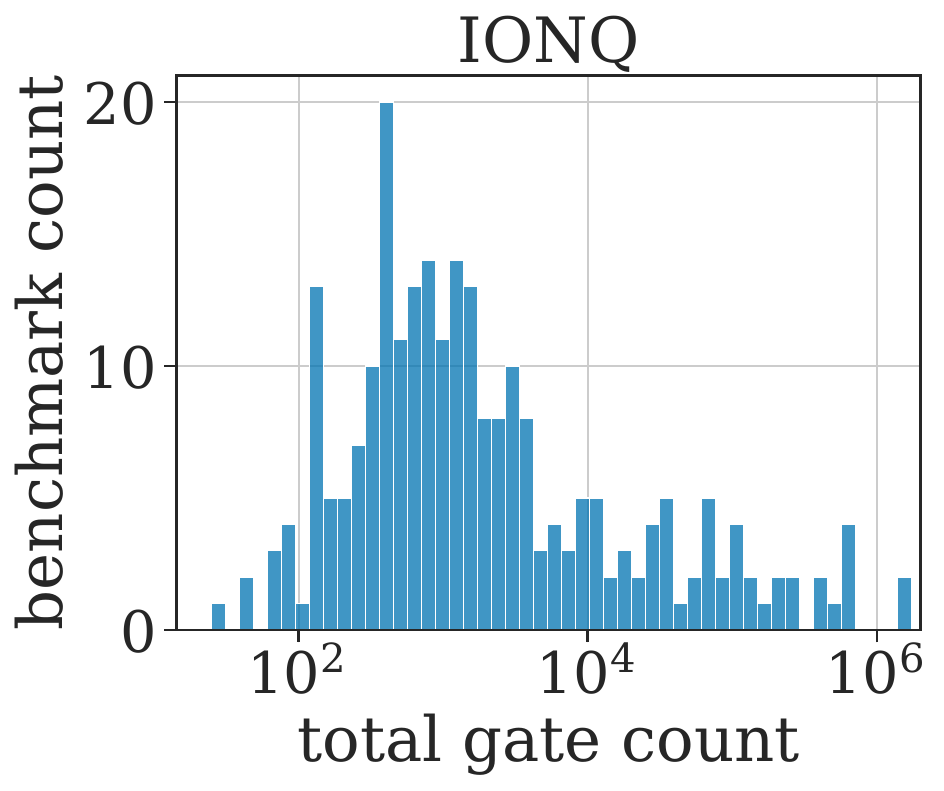}
        \end{subfigure}
    \caption{Summary of the benchmarks' original total gate counts across all gate sets (log-scale $x$-axis).} 
    \label{fig:benchmark-sizes}
\end{figure}

\bibliographystyle{plain}
\bibliography{references}

\end{document}